\documentclass[twocolumn,showpacs,pra,amsmath,amssymb]{revtex4}
\usepackage{bm}
\usepackage{graphicx}
\newcommand{\bfsfI}{\textsf{\textbf{I}}}
\newcommand{\bfsfG}{\textsf{\textbf{G}}}
\newcommand{\bfsff}{\textsf{\textbf{f}}}
\newcommand{\half}{{\textstyle \frac{1}{2}}}
\newcommand{\threehalf}{{\textstyle \frac{3}{2}}}
\newcommand{\fivehalf}{{\textstyle \frac{5}{2}}}
\newcommand{\fivefourth}{{\textstyle \frac{5}{4}}}
\newcommand{\fourthird}{{\textstyle \frac{4}{3}}}
\newcommand{\shortdash}{\protect\rule[0.5mm]{0.7mm}{0.1mm}}
\newcommand{\longdash}{\protect\rule[0.5mm]{2mm}{0.1mm}}
\newcommand{\points}{\protect\rule[0.5mm]{0.2mm}{0.2mm}}

\begin{document}
\title{Modified atomic decay rate near absorptive scatterers at finite temperature}
\date{\today
}
\author{L.G.~Suttorp}\email{l.g.suttorp@uva.nl}
\author{A.J.~van Wonderen} 
\affiliation{Institute for Theoretical Physics, University of Amsterdam,
Science Park 904, 1098 XH Amsterdam, The Netherlands}

\begin{abstract}
  The change in the decay rate of an excited atom that is brought about by
  extinction and thermal-radiation effects in a nearby dielectric medium is
  determined from a quantummechanical model. The medium is a collection of
  randomly distributed thermally-excited spherical scatterers with
  absorptive properties. The modification of the decay rate is described by
  a set of correction functions for which analytical expressions are
  obtained as sums over contributions from the multipole moments of the
  scatterers. The results for the modified decay rate as a function of the
  distance between the excited atom and the dielectric medium show the
  influence of absorption, scattering and thermal-radiation processes. Some
  of these processes are found to be mutually counteractive. The changes in
  the decay rate are compared to those following from an effective-medium
  theory in which the discrete scatterers are replaced by a continuum.
\end{abstract}  
\maketitle

\section{Introduction\label{sectionintro}}

The emission process of a photon by an excited atom is influenced by its
environment \cite{P46,D70,CPS74}.  In particular, when the excited atom is
located in the vicinity of a dielectric body, its decay rate will be
different from that in vacuum. It will depend crucially on the precise
properties of the dielectric medium for several reasons. Interference
effects due to the reflection of emitted photons from the surface of a
dielectric medium will alter the atomic decay rate. Furthermore, if the
medium is dispersive and lossy, it may absorb photons. This absorption
process will lead to an enhanced atomic decay. Scattering effects due to
inhomogeneities in the dielectric medium will also change the atomic
decay. Finally, if the dielectric has got a finite temperature, it will
emit thermal radiation which will stimulate the emission process. All these
effects will depend on the shape of the dielectric body and on its distance
from the atom.

Detailed studies of the changes in the atomic decay rate due to the
presence of homogeneous lossy dielectric media at zero temperature have
been carried out for various geometries. The simplest case is that of a
halfspace that is uniformly filled with a dispersive dielectric
\cite{A75,YG96,SKW99,EZ12}. A more complicated configuration is that of a
dielectric sphere, which has been studied extensively
\cite{R82,AON83,KLG88,DKW00,DKW01,KL05,CLT12a,CLT12b}.  Other geometries
like cylinders or spheroids have been considered as well \cite{KDL01}.

In a realistic medium extinction of radiation is produced not only by
absorption but also by scattering from inhomogeneities. Both these effects
will alter the decay rate of a nearby atom. When the inhomogeneities are
densely distributed, multiple-scattering effects greatly complicate the
analysis. For such systems numerical simulations have been used to gain
insight in modified emission processes \cite{YV07a,YV07b,FPCS07,PC10}.  For
a dilute distribution of scatterers, however, multiple scattering plays a
minor role, so that analytical methods may be used. In a recent paper
\cite{SvW11} we have studied a model in which an excited atom is situated
near a halfspace that is filled with a dilute set of spherical scatterers
consisting of absorptive dielectric material.  For this model detailed
results for the change in the atomic decay rate as a function of the
distance between atom and halfspace have been obtained.

The changes in the atomic decay rate due to thermal radiation from a nearby
dielectric medium are of a different nature from those discussed above, as
they are a consequence of stimulated-emission effects. Although their
origin is different, they are likewise sensitive to the geometric details
of the dielectric medium. For a uniform medium with a flat interface
thermal-radiation effects in the atomic emission process have been studied
in \cite{SJCG00,JMMCG05}.

In the present paper we shall treat within a single model decay-rate
changes due to both extinction effects (from absorption and scattering) and
to thermal radiation. We shall consider a randomly distributed dilute set
of spherical scatterers consisting of a lossy dielectric material. By
choosing the scatterers to be spherical as in \cite{SvW11} we are able to
use the framework provided by Mie theory. Taking the set of scatterers to
be dilute permits us to neglect multiple-scattering effects, as explained
above. The scatterers will be held at a finite temperature so that they
will emit thermal radiation. We will choose the set of scatterers to be
bounded, so that the optical depth of the aggregate stays finite. For
analytical purposes, the shape of the aggregate will be taken to be
spherical. The model will allow us to treat the effects of reflective
interference, absorption, scattering and thermal radiation on an equal
footing.

The paper is organized as follows. In Section \ref{sectiondecay} we
introduce our model that is based on the damped-polariton formulation of
absorptive dielectrics. We formulate the atomic decay rate in terms of the
second moment of the electric field and the associated Green function. In
Section \ref{sectioncoldhot} we use addition theorems for vectorial
spherical wave functions to evaluate the correction functions that govern
the changes in the atomic decay rate in the presence of a set of absorptive
scatterers at finite temperature. Subsequently, an alternative approach to
the correction functions is given in Section \ref{sectionintegral}. It
makes use of integral representations, which are more suitable for
numerical evaluation. Some typical examples of the ensuing graphs for the
correction functions are presented in Section
\ref{sectionevaluation}. Furthermore, a comparison is made with an
effective-medium model in which a uniform dielectric replaces the set of
discrete scatterers. In a final Section \ref{sectiondiscussion} our results
are summarized and discussed. A few technical details of our treatment are
given in a set of Appendices.

\section{Atomic emission and absorption near a dispersive 
dielectric medium\label{sectiondecay}}

A dispersive and absorptive linear dielectric medium may be described by a
damped-polariton model that has been introduced some time ago \cite{HB92}.
In this model damping effects are taken into account by coupling the
polarization density to a bath of harmonic oscillators with a continuous
range of frequencies. For a uniform dielectric medium Fourier expansions
can be used to diagonalize the model Hamiltonian and to determine its
eigenmodes. For arbitrarily inhomogeneous media diagonalization can
likewise be carried out by means of a Green function technique
\cite{SvW04}. In the latter case diagonalization leads to a Hamiltonian of
the form
\begin{equation}
H=\int d  \mathbf{r}\int_0^{\infty} d \omega\, \hbar\omega \, 
\mathbf{C}^{\dagger}(\mathbf{r},\omega)\cdot \mathbf{C}(\mathbf{r},\omega)\, , \label{1}
\end{equation}
with annihilation operators $\mathbf{C}(\mathbf{r},\omega)$ and associated
creation operators depending on position and frequency arguments. These
fulfill the standard commutation relations $[\mathbf{C}(\mathbf{r},\omega),
\mathbf{C}^{\dagger}(\mathbf{r}',\omega')]=\bfsfI\, \delta(\mathbf{r}-
\mathbf{r}')\delta(\omega-\omega')$, with $\bfsfI$ the unit tensor. The
electric field can be expressed in terms of these operators as $\mathbf{E}
(\mathbf{r})=\int_0^{\infty} d\omega \,
\mathbf{E}(\mathbf{r},\omega)+\mathrm{h.c.}$, with
\begin{equation}
\mathbf{E}(\mathbf{r},\omega)=\int d\mathbf{r}'\,
\bfsff_E(\mathbf{r},\mathbf{r}',\omega)\cdot\mathbf{C}(\mathbf{r}',\omega)\, .\label{2} 
\end{equation} 
The tensorial coefficient $\bfsff_E$ reads \cite{SvW04}:
\begin{equation}
\bfsff_E(\mathbf{r},\mathbf{r}',\omega)=-i\, \frac{\omega^2}{c^2}
\left[\frac{\hbar \, \mathrm{Im}\,\varepsilon
(\mathbf{r}',\omega+i0)}{\pi\varepsilon_0}\right]^{1/2}\, 
\bfsfG(\mathbf{r},\mathbf{r}',\omega+i 0) \, . \label{3}
\end{equation} 
Here $\varepsilon$ is the complex local (relative) dielectric constant,
which follows from the parameters of the model. Furthermore, $\bfsfG$ is
the tensorial Green function, which satisfies the differential equation
\begin{eqnarray}
 && -\nabla\times [\nabla\times \bfsfG (\mathbf{r},
\mathbf{r}',\omega+i 0)]\nonumber\\
&&+\frac{\omega^2}{c^2}\, \varepsilon
(\mathbf{r},\omega+i 0)\, 
\bfsfG(\mathbf{r},\mathbf{r}',\omega+i 0)
=\bfsfI\, \delta(\mathbf{r}-\mathbf{r}')\, .
 \label{4}
\end{eqnarray}

The rate of photon emission by an excited atom in the vicinity of a
dispersive and absorptive dielectric follows from the inhomogeneous
damped-polariton model in its diagonalized form by employing perturbation
theory in leading order and disregarding transient effects \cite{SvW10}. In
the electric-dipole approximation it can be expressed as an integral over
the second moment of the electric field:
\begin{eqnarray}
 \Gamma=\frac{2\pi}{\hbar^2}
\langle e|{\bm \mu}|g\rangle \cdot
\int_0^\infty d \omega' \, 
\langle \mathbf{E}(\mathbf{r}_a,\omega)\mathbf{E}^\dagger(\mathbf{r}_a,\omega')\rangle
\cdot\langle g|{\bm \mu}|e\rangle\, , &&
\nonumber\\
\label{5}
\end{eqnarray}
with atomic dipole matrix elements connecting the excited state (labelled
$e$) and the ground state ($g$) of the atom, with $\omega$ the atomic
transition frequency and with $\mathbf{r}_a$ the atomic position. The
second moment of the electric field is determined by the density matrix of
the damped-polariton system.

After insertion of (\ref{2}) one encounters the second moment of the
creation and annihilation operators. This second moment will be assumed to
be isotropic and diagonal in both the position and the frequency variables:
\begin{equation}
\langle \mathbf{C}(\mathbf{r},\omega) \, \mathbf{C}^\dagger(\mathbf{r}',\omega')\rangle=
[1+g(\mathbf{ r},\omega)]\, \bfsfI \, \delta(\mathbf{r}-\mathbf{r}')\delta(\omega-\omega')\, ,
\label{6}
\end{equation}
with a real and positive function $g(\mathbf{r},\omega)$ that vanishes when
the damped-polariton system, as given by the Hamiltonian (\ref{1}), is in
its ground state. The second moment of the electric field now gets the form
\begin{eqnarray}
&&\langle \mathbf{E}(\mathbf{r},\omega) \mathbf{E}^\dagger(\mathbf{r},\omega')\rangle=
\delta(\omega -\omega')\frac{\hbar\omega^4}{\pi\varepsilon_0 c^4}\nonumber\\
&&\times\int d\mathbf{r}' \, \bfsfG(\mathbf{r},\mathbf{r}',\omega+i0)\cdot 
\tilde{\bfsfG}^\ast(\mathbf{r},\mathbf{r}',\omega+i0) \nonumber\\
&&\times [1+g(\mathbf{r}',\omega)]\, \mathrm{Im} \,\varepsilon(\mathbf{r}',\omega+i0)\, ,
\label{7}
\end{eqnarray}
with $\tilde{\bfsfG}$ the transpose of $\bfsfG$. The part of the integral
that is independent of $g(\mathbf{r},\omega)$ can be rewritten with the
help of the optical theorem for the Green function:
\begin{eqnarray}
&&\int d\mathbf{r}'  \, \bfsfG(\mathbf{r},\mathbf{r}',\omega+i0)\cdot 
\tilde{\bfsfG}^\ast(\mathbf{r},\mathbf{r}',\omega+i0) \, \mathrm{Im}\, 
\varepsilon(\mathbf{r}',\omega+i0)=\nonumber\\
&&=-\frac{c^2}{\omega^2}\, \mathrm{Im}\, \bfsfG(\mathbf{r},\mathbf{r},\omega+i0)\, .
\label{8}
\end{eqnarray}

To evaluate the second part of the integral in (\ref{7}) one needs
information on the function $g(\mathbf{r}',\omega)$ in the integrand. If
$g(\mathbf{r}',\omega)$ stays finite for large $|\mathbf{r}|'$, the distant
regions in the integral may contribute even when $\mathrm{Im}\,
\varepsilon$ tends to 0 there. The importance of the far parts of space
while evaluating the second moment of the electric field has been discussed
previously \cite{HR09}.  In fact, the Green function
$\bfsfG(\mathbf{r},\mathbf{r}',\omega+i0)$ for a system with a uniform
dielectric constant $\varepsilon$ at large distances from the origin will
be proportional to $e^{i\omega\varepsilon|\mathbf{r}'|/c}/ |\mathbf{r}'|$
for large $|\mathbf{r}'|$, so that upon integration these distant regions
yield a finite contribution even when $\mathrm{Im}\, \varepsilon$ gets
negligibly small. However, if $g(\mathbf{r}',\omega)$ tends to 0 for large
$|\mathbf{r}'|$, as is reasonable when no incoming fields are present, the
contribution of the far spatial domain to the integral is no longer
important when $\mathrm{Im}\, \varepsilon$ gets small there. In that case
the integration in the second part of the integral in (\ref{7}) can be
confined to positions in space where $\mathrm{Im}\,\varepsilon$ differs
from zero by a physically relevant amount, as will be indicated by a
subscript $V$ at the integral. Upon assuming moreover for simplicity that
both $g(\mathbf{r}',\omega)$ and $\mathrm{Im}\,
\varepsilon(\mathbf{r}',\omega+i0)$ are independent of position within $V$,
we arrive at the following expression for the second moment of the electric
field:
\begin{eqnarray}
&&\langle \mathbf{E}(\mathbf{r},\omega) \mathbf{E}^\dagger(\mathbf{r},\omega')\rangle=
-\delta(\omega-\omega')\frac{\hbar\omega^2}{\pi\varepsilon_0 c^2}\, \mathrm{Im} \,
\bfsfG(\mathbf{r},\mathbf{r},\omega+i0)\nonumber\\
&&+\delta(\omega-\omega')\frac{\hbar\omega^4}{\pi\varepsilon_0
  c^4}g(\omega)\, \left[\mathrm{Im}\, \varepsilon(\omega+i0)\right]\nonumber\\
&&\times \int_V d\mathbf{r}'
\bfsfG(\mathbf{r},\mathbf{r}',\omega+i0)\cdot 
\tilde{\bfsfG}^\ast(\mathbf{r},\mathbf{r}',\omega+i0) \, .
\label{9}
\end{eqnarray}
For a dielectric at a finite inverse temperature $\beta$ one may insert
$g(\omega)=1/(e^{\beta\hbar\omega}-1)$.  

The atomic decay rate follows by substituting (\ref{9}) in (\ref{5}).  As a
result the decay rate is found as the sum of two terms, one representing
spontaneous emission in the presence of an absorbing dielectric at zero
temperature, and the other a correction describing stimulated emission due
to the finite-temperature radiation from the dielectric. These two
contributions have different properties and should be treated separately,
as has been remarked before \cite{SJCG00,SSG02,HR09}.

In the presence of thermal radiation from the dielectric, atomic
transitions from the ground state to the excited state may occur as
well. The ensuing rate of atomic absorption is determined by an expression
similar to (\ref{5}):
\begin{eqnarray}
 \Gamma_{\mathrm{a}}=\frac{2\pi}{\hbar^2}
\langle g|{\bm \mu}|e\rangle \cdot
\int_0^\infty d \omega' \, 
\langle \mathbf{E}^\dagger(\mathbf{r}_a,\omega)\mathbf{E}(\mathbf{r}_a,\omega')\rangle
\cdot\langle e|{\bm \mu}|g\rangle\, . &&
\nonumber\\
\label{10}
\end{eqnarray}
The second moment of the electric field occurring here is given by the
analogue of (\ref{9}), with no first term and with a second term that
follows by interchanging $\bfsfG$ and $\bfsfG^\ast$.

\section{Decay near a sphere with absorptive
  scatterers\label{sectioncoldhot}}

We consider a spherical domain with radius $R$ that contains a dilute set
of non-overlapping spherical scatterers with radii $a\ll R$ and complex
dielectric constant $\varepsilon$, while the space between the scatterers
is empty vacuum. The region $V$ in (\ref{9}) is given by the union of the
interiors of the spherical scatterers. The enveloping spherical domain with
volume $4\pi R^3/3$ will be indicated by $V_0$. The scatterers are assumed
to be randomly distributed with a uniform average density. An excited atom
is located outside the spherical domain, at a distance $r_a>R+a$ from its
center, which is chosen as the origin of a spherical coordinate system. We
wish to determine the change in the decay rate that is brought about by the
presence of the set of spherical scatterers.  In this section we will first
concentrate on the effects of scattering for the case of cold
scatterers. Subsequently, finite-temperature effects will be considered as
well.

For cold scatterers the second moment of the electric field is given by the
first term of (\ref{9}).  For a single cold scatterer the Green function
$\bfsfG$ is the sum of a vacuum contribution $\bfsfG_0$ and a scattering
term $\bfsfG_\mathrm{s}$, as given in Appendix \ref{appendixa}. The
imaginary part of the vacuum Green function with coinciding position
arguments is proportional to the unit tensor: $\mathrm{Im}
\,\bfsfG_0(\mathbf{r},\mathbf{r},\omega+i0)=-\omega\bfsfI/(6\pi c)$,
whereas the imaginary part of the scattering term $\bfsfG_\mathrm{s}$ for
coinciding positions follows from (\ref{A1}).

For a dilute set of $N_s$ scatterers, with centers at the positions
$\mathbf{r}_i$, the Green function can be approximated as
\begin{eqnarray}
&&\bfsfG(\mathbf{r},\mathbf{r}',\omega+i0)=\bfsfG_0(\mathbf{r},\mathbf{r}',\omega+i0)\nonumber\\
&&+\sum_i \bfsfG_\mathrm{s}(\mathbf{ r}-\mathbf{r}_i,\mathbf{r}'-\mathbf{r}_i,\omega+i0)\, ,
\label{11}
\end{eqnarray}
as follows by suppressing all terms involving multi-scatterer
configurations in the Foldy-Lax equations or, more directly, in the
Neumann series for the integral equation that determines the Green function
\cite{TKS85,TK01}.  Adopting this form of the Green function implies that
multiple-scattering effects are henceforth neglected, as is reasonable for a
dilute set.  Since the centers of the scattering spheres are located at
random positions inside the spherical domain with radius $R$, the
configurational average of the second term of (\ref{11}) may be written as
an integral over $V_0$:
\begin{equation}
n_s\int_{V_0} d\mathbf{r}'' \bfsfG_\mathrm{s}(\mathbf{ r}-\mathbf{r}'',\mathbf{r}'-
\mathbf{r}'',\omega+i0)
\, ,\label{12}
\end{equation} 
with $n_s=3N_s/(4\pi R^3)$ the density of the scatterers. Upon inserting
the expression (\ref{A1}) for $\bfsfG_\mathrm{s}$, employing the addition
theorem (\ref{A6}) for the vector spherical wave functions
$\mathbf{M}^{(h)}_{l,m}(\mathbf{r})$, $\mathbf{N}^{(h)}_{l,m}(\mathbf{r})$
and carrying out the integral over $\mathbf{r}''$, we get for the imaginary
part of the configurationally averaged Green function with coinciding
position arguments:
\begin{eqnarray}
&&\mathrm{Im}\, \langle\bfsfG(\mathbf{r},\mathbf{r},\omega+i0)\rangle=
-\frac{\omega}{6\pi c}\bfsfI+4\pi \frac{\omega}{c} n_s \sum_{l=1}^{\infty}
l(l+1)\nonumber\\
&&\times\sum_{l'=1}^\infty\sum_{m'=-l'}^{l'}\sum_{l''=0}^\infty
(-1)^{m'}\frac{2l''+1}{l'(l'+1)} I_{l''}(R) 
\left(\begin{array}{ccc}
l & l' & l''\\
1 & -1 & 0\end{array}\right)^2\nonumber\\
&&\times\mathrm{Re}\left[ (-i)^{l+1} B^e_l\left\{ 
\delta^\mathrm{e}_{l+l'+l''}
\mathbf{N}^{(h)}_{l',m'}(\mathbf{r})\mathbf{N}^{(h)}_{l',-m'}
(\mathbf{r})\right.\right.\nonumber\\
&&\left.+\delta^\mathrm{o}_{l+l'+l''}
\mathbf{M}^{(h)}_{l',m'}(\mathbf{r})\mathbf{M}^{(h)}_{l',-m'}(\mathbf{r})\right\}\nonumber\\
&&+(-i)^{l+1} B^m_l\left\{ 
\delta^\mathrm{e}_{l+l'+l''}
\mathbf{M}^{(h)}_{l',m'}(\mathbf{r})\mathbf{M}^{(h)}_{l',-m'}(\mathbf{r})\right.\nonumber\\
&&\left.\left.+\delta^\mathrm{o}_{l+l'+l''}
\mathbf{N}^{(h)}_{l',m'}(\mathbf{r})\mathbf{N}^{(h)}_{l',-m'}(\mathbf{r})\right\}
\right]\, .
\label{13}
\end{eqnarray}
Here $I_l(R)=\int_0^R dr\, r^2[j_l(kr)]^2$, with $k=\omega/c$, results from
the integration over the spherical domain. Its explicit form is
\begin{eqnarray}
I_l(R)&=&\half
R^3[j_l(kR)]^2+\half R^3 [j_{l+1}(kR)]^2\nonumber\\
&&-\half (2l+1)(R^2/k)j_l(kR)j_{l+1}(kR)\, ,
\label{14}
\end{eqnarray}
as may be checked by differentiating with respect to $R$ and employing the
recursion relations for the spherical Bessel functions \cite{NIST10}. The
Wigner 3$j$-symbols in (\ref{13}) imply that the three summation variables $l,l',l''$
satisfy the triangular conditions $|l'-l''|\leq l \leq l'+l''$. Furthermore,
$\delta^\mathrm{e}_l$ equals 1 for even $l$, and 0 for odd $l$, while
$\delta^\mathrm{o}_l$ is defined analogously, with even and odd
interchanged.

Substitution of (\ref{13}) in the first term of (\ref{9}) yields an
expression for the configurational average of the second moment of the
electric field. From the spherical symmetry it follows that the resulting
tensor is diagonal in a spherical coordinate system with origin at the
center of the aggregate of scatterers. Upon using the expressions for the
vector spherical wave functions in spherical coordinates \cite{FJ87} one
gets
\begin{eqnarray}
&&\langle\!\langle\mathbf{E}(\mathbf{r},\omega)
\mathbf{E}^\dagger
(\mathbf{r},\omega')\rangle\!\rangle=
\delta(\omega-\omega')\frac{\hbar \omega^3}{6\pi^2\varepsilon_0 c^3}\nonumber\\
&&\times\left\{\left[1+f F_{\mathrm{c},\parallel}(r)\right]\, \mathbf{e}_r\mathbf{e}_r
+\left[1+f F_{\mathrm{c},\perp}(r)\right]\, 
\left(\mathbf{e}_\theta\mathbf{e}_\theta+\mathbf{e}_\varphi\mathbf{e}_\varphi\right)\right\}\, ,\nonumber\\
\label{15}
\end{eqnarray}
with the combined configurational and density-matrix average indicated by
double brackets. Furthermore, $\mathbf{e}_r$, $\mathbf{e}_\theta$ and
$\mathbf{e}_\varphi$ are unit vectors in spherical coordinates, and
$f=\fourthird \pi a^3 n_s$ is the filling fraction of the spherical domain containing
the scatterers. The
decay-rate correction functions $F_{\mathrm{c},\parallel}$ and
$F_{\mathrm{c},\perp}$ are
\begin{eqnarray}
&&F_{\mathrm{c},\parallel}(r)=-\frac{9}{2a^3}\sum_{l=1}^\infty l(l+1)\sum_{l'=1}^\infty\sum_{l''=0}^\infty
 l'(l'+1)c_{l,l'.l''}(R)\nonumber\\
&&\times\mathrm{Re}\left[
(-i)^{l+1}\left(B^e_l
    \delta^\mathrm{e}_{l+l'+l''}+B^m_l\delta^\mathrm{o}_{l+l'+l''}\right)
\frac{1}{k^2r^2}\, H_{l'}(kr)
\right]\, ,\nonumber\\
\label{16}
\end{eqnarray} 
\begin{eqnarray}
&&F_{\mathrm{c},\perp}(r)=-\frac{9}{4a^3}\sum_{l=1}^\infty l(l+1)\sum_{l'=1}^\infty\sum_{l''=0}^\infty
c_{l,l'.l''}(R)\nonumber\\
&&\times
\mathrm{Re}
\biggl[ (-i)^{l+1}\left(B^e_l
    \delta^\mathrm{e}_{l+l'+l''}
+B^m_l\delta^\mathrm{o}_{l+l'+l''}\right)
\frac{1}{k^2r^2}\, \bar{H}_{l'}(kr)
\biggr.
\nonumber\\
\nonumber\\
&&\left.+(-i)^{l+1}\left(B^e_l\delta^\mathrm{o}_{l+l'+l''}+B^m_l\delta^\mathrm{e}_{l+l'+l''}\right)
H_{l'}(kr)
\right]
\, , \nonumber\\
\label{17}
\end{eqnarray}
with the functions $H_l(t)=[h^{(1)}_l(t)]^2$ and
$\bar{H}_l(t)=[d[th^{(1)}_l(t)]/dt]^2$, and with the coefficients
\begin{equation}
c_{l,l',l''}(R)=(2l'+1)(2l''+1)I_{l''}(R)
\left(\begin{array}{ccc}
l & l' & l''\\
1 & -1 & 0\end{array}\right)^2 \, .
\label{18}
\end{equation}

In view of (\ref{5}) the average decay rate for an excited atom in the
presence of a randomly distributed set of cold spherical scatterers can be
written as
\begin{equation}
\langle\Gamma_\mathrm{c}\rangle=\Gamma_0+f\left[F_{\mathrm{c},\parallel}(r_a)\Gamma_{0,\parallel}+
F_{\mathrm{c},\perp}(r_a)\Gamma_{0,\perp}\right]\, ,
\label{19}
\end{equation}
with the vacuum decay rate $\Gamma_0=\Gamma_{0,\parallel}+\Gamma_{0,\perp}$
for $\Gamma_{0,\parallel}=\omega^3|\langle e|{\bm
  \mu}\cdot\mathbf{e}_r|g\rangle|^2/(3\pi\varepsilon_0\hbar c^3)$ and
$\Gamma_{0,\perp}=\omega^3(|\langle e|{\bm
  \mu}\cdot\mathbf{e}_\theta|g\rangle|^2+|\langle e|{\bm
  \mu}\cdot\mathbf{e}_\varphi|g\rangle|^2)/(3\pi\varepsilon_0\hbar c^3)$.
This expression for the decay rate in the presence of an aggregate of cold
scatterers reduces to a simpler form if only a single scatterer is
present. The latter form is found by taking the limit $R\rightarrow 0$ and
$n_s\rightarrow \infty$ with $4\pi R^3 n_s/3=1$, so that $n_s\,
c_{l,l',l''}(R)$ gets equal to $\delta_{l,l'}\delta_{l'',0}/(4\pi)$. As a
consequence, only the terms with $\delta^\mathrm{e}_{l+l'+l''}$ in the
correction functions survive in this limit. The resulting expressions agree
with those given in \cite{AON83,KLG88,DKW01}.

For the general case involving a collection of scatterers the decay-rate
correction functions $F_{\mathrm{c},\parallel}$ and $F_{\mathrm{c},\perp}$
show a symmetry in the contributions of the electric and magnetic multipole
amplitudes. This symmetry gets lost for a single scatterer.  Both
correction functions depend on the distance $r_a$ of the atom from the
center of the spherical domain containing the scatterers, and parametrically on the
radius $R$ of the spherical domain and on the radii $a$ of the scatterers
themselves (through the scattering amplitudes $B^e_l$ and $B^m_l$). These
three independent length scales that together characterize the
configuration (all measured in terms of the wavelength of the atomic
transition) occur neatly separated in the summands in (\ref{16}) and
(\ref{17}). In particular, the dependence on $r_a$ is given by the
spherical Hankel functions $h^{(1)}_{l'}(kr_a)$ and their
derivatives. Since these are proportional to $e^{ikr_a}$, the correction
functions will be oscillating as a function of $r_a$ (on the scale of the
wavelength). For large $r_a$ the correction functions will decay to 0.

When the scatterers are taken to be at finite temperature the second moment
of the electric field gets an additional contribution that is given by the
second term in (\ref{9}). The integral over $V$ in that term is taken over
all positions $\mathbf{r}'$ inside the scatterers, so that it is in fact a
sum over individual contributions for each of the scatterers. As the system
of scatterers is dilute, one may use the expression (\ref{A9}) for the
Green functions in the integrand. In this way the hot-scatterer
contribution to the second moment becomes
\begin{eqnarray}
&&\delta(\omega-\omega')\frac{\hbar
  \omega^4}{\pi\varepsilon_0 c^4(e^{\beta\hbar\omega}-1)}
\left[\mathrm{Im}\,\varepsilon(\omega+i0)\right]\sum_i \int_{|\mathbf{r}'-\mathbf{r}_i|<a}
d\mathbf{r}' \nonumber\\
&&\times\bfsfG_{0\mathrm{s}}(\mathbf{r}-\mathbf{r}_i,\mathbf{r}'-\mathbf{r}_i,\omega+i0)\cdot
\tilde{\bfsfG}^\ast_{0\mathrm{s}}(\mathbf{r}-\mathbf{r}_i,\mathbf{r}'-\mathbf{r}_i,\omega+i0)\, .
\nonumber\\
&&
\label{20}
\end{eqnarray}
When the explicit expression (\ref{A9}) for $\bfsfG_{0\mathrm{s}}$ is
substituted, the integration over $\mathbf{r}'$ leads to three-dimensional
integrals of scalar products of vector spherical wave functions of the form
(\ref{A11})--(\ref{A13}). The resulting integrals $I^{(\varepsilon)}_l(a)$
may be eliminated with the help of (\ref{A15})--(\ref{A16}). Furthermore,
the sum over $i$ may be replaced by an integral over positions
$\mathbf{r}''$ inside the spherical domain of radius R.  In this way we get
from (\ref{20}):
\begin{eqnarray}
&&\delta(\omega-\omega')\frac{\hbar\omega^3}{\pi\varepsilon_0
  c^3(e^{\beta\hbar\omega}-1)}
n_s\int_{V_0} d\mathbf{r}'' \sum_{l=1}^\infty\sum_{m=-l}^l \frac{1}{2l+1}\nonumber\\
&&\times\left[ C^m_l \mathbf{M}^{(h)}_{l,m}(\mathbf{r}-\mathbf{r}'')
\mathbf{M}^{(h)\ast}_{l,m}(\mathbf{r}-\mathbf{r}'')\right. \nonumber\\
&&\left.+C^e_l \mathbf{N}^{(h)}_{l,m}(\mathbf{r}-\mathbf{r}'')
\mathbf{N}^{(h)\ast}_{l,m}(\mathbf{r}-\mathbf{r}'')\right]\, . \label{21}
\end{eqnarray}
The remaining vector spherical wave functions, which depend on
$\mathbf{r}-\mathbf{r}''$, can be rewritten by using the addition theorem
of Appendix \ref{appendixa}. Finally, we arrive at an expression for the
second moment that is a generalization of (\ref{15}), with the
cold-scatterer decay-rate corrections functions $F_{\mathrm{c},p}(r)$ (with
$p=\, \parallel,\perp$) replaced by new functions $F_p(r)$. These contain,
apart from the cold-scatterer correction functions, additional terms that
arise when the dielectric medium in the scatterers emits thermal
radiation. In fact, they have the form
$F_{p}(r)=F_{\mathrm{c},p}(r)+(e^{\beta\hbar\omega}-1)^{-1}\,
F_{\mathrm{d},p}(r)$, with the `dielectric' decay-rate correction functions
$F_{\mathrm{d},\parallel}(r)$ and $F_{\mathrm{d},\perp}(r)$.  (The
subscript $\mathrm{d}$ points to the fact that the hot-scatterer
contributions find their origin in the dielectric medium.) The dielectric
decay-rate correction functions have a similar form as (\ref{16}) and
(\ref{17}), with the following differences: 1. the functions $H_l(t)$ and
$\bar{H}_l(t)$ are replaced by $H_{\mathrm{d},l}(t)=|h^{(1)}_l(t)|^2$ and
$\bar{H}_{\mathrm{d},l}(t)=|d[th^{(1)}_l(t)]/dt|^2$, respectively; 2. the
multipole amplitudes $(-i)^{l+1}\, B^q_l$ (with $q=e,m$) are replaced by
$-C^q_l$, defined in (\ref{A17}); 3. the symbol $\mathrm{Re}$ can be
omitted (after the above changes the functions are real).

The average decay rate $\langle\Gamma\rangle$ for an excited atom in the
presence of an aggregate of randomly distributed spherical scatterers at
finite temperature gets the form:
\begin{equation}
\langle\Gamma\rangle=\Gamma_0+f\left[F_{\parallel}(r_a)\Gamma_{0,\parallel}+
F_{\perp}(r_a)\Gamma_{0,\perp}\right]\, ,
\label{22}
\end{equation}
instead of (\ref{19}). As we have seen in Section \ref{sectiondecay}, the
cold-scatterer contribution to the average decay rate follows from the
first term in (\ref{9}), which resulted by rewriting part of the integral
in (\ref{7}) by means of the optical theorem (\ref{8}). Hence, it
originates from positions $\mathbf{r}'$ both in the scatterers and in the
surrounding space. In contrast, the hot-scatterer contribution has been
obtained from the second term in (\ref{9}), which is an integral over the
volume $V$ occupied by the dielectric medium within the scatterers. To show
more clearly the origin of the various contributions we may rearrange the
total decay-rate correction functions $F_p$ by writing $F_{\mathrm{c},p}$
as the sum $F_{\mathrm{d},p}+F_{\mathrm{r},p}$ of the dielectric decay-rate
correction function $F_{\mathrm{d},p}$ and a `radiative' decay-rate
correction function $F_{\mathrm{r},p}=F_{\mathrm{c},p}-F_{\mathrm{d},p}$,
so that one gets
\begin{equation}
F_p(r)=
F_{\mathrm{r},p}(r)+\frac{e^{\beta\hbar\omega}}{e^{\beta\hbar\omega}-1} 
F_{\mathrm{d},p}(r)\, ,
\label{23}
\end{equation}
with $p=\parallel,\perp$.  In this form the last term contains all
contributions from the dielectric medium within the scatterers (both for
cold and for hot scatterers), whereas the first term represents radiative
contributions associated to the space outside the scatterers.

The average absorption rate $\langle\Gamma_\mathrm{a}\rangle$ for a
ground-state atom in the vicinity of a collection of randomly distributed
hot scatterers can likewise be evaluated. As absorption and stimulated
emission are closely related, it is found to be determined by the same
correction functions $F_{\mathrm{d},\parallel}$ and $F_{\mathrm{d},
  \perp}$:
\begin{equation}
\langle\Gamma_{\mathrm{a}}\rangle=\frac{f}{e^{\beta\hbar\omega}-1}
\left[F_{\mathrm{d},\parallel}(r_a)\Gamma_{0,\parallel}+
F_{\mathrm{d},\perp}(r_a)\Gamma_{0,\perp}\right]\, .
\label{24}
\end{equation}
 
The dielectric decay-rate correction functions for hot scatterers are quite
analogous to their cold-scatterer counterparts (\ref{16}) and
(\ref{17}). In particular, the symmetry between electric and magnetic
multipole contributions is clearly visible in these functions as well. It
gets lost in the limiting case of a single hot scatterer. As before, the
independent length scales $r_a$, $R$ and $a$ show up in the summand. The
dependence on $a$ is contained in the coefficients $C^e_l$ and $C^m_l$,
while $R$ appears in the coefficients (\ref{18}). The distance $r_a$ enters
the summand through the absolute value of the spherical Hankel
functions. As a consequence, the oscillating behavior found before is
absent here. In fact, the dielectric correction functions will decay
smoothly to 0 when the distance $r_a$ increases beyond bounds. In the next
section we shall determine the precise form of the asymptotic behavior for
all decay-rate correction functions.

\section{Integral representations for the decay-rate correction
  functions}\label{sectionintegral}

The expressions for the decay-rate correction functions as found in the
previous section are three-fold sums of terms in which the independent
variables $r_a$, $R$ and $a$ occur in separate factors. To determine their
behavior as a function of $r_a$, for fixed values of the parameters $R$ and
$a$, the sums have to be evaluated numerically. As it turns out, a greater
numerical efficiency is achieved by starting from an alternative
representation in which the variables $r_a$ and $R$ are intertwined in
integral expressions. For cold scatterers such a representation can be
found by starting again from the first term of (\ref{9}) and using
(\ref{11})-(\ref{12}), as before. Upon choosing the atomic position
$\mathbf{r}_a=\mathbf{r}=\mathbf{r}'$ on the positive $z$-axis and
inserting the representation (\ref{A8}), one may write the $zz$-component
of the integrand in (\ref{12}) as $\cos^2\chi\,
G_{\mathrm{s},\parallel}(t,\omega+i0)+\sin^2\chi \,
G_{\mathrm{s},\perp}(t,\omega+i0)$, with $t=|\mathbf{r}-\mathbf{r}''|$ and
$\chi$ the angle between $\mathbf{r}-\mathbf{r}''$ and the positive
$z$-axis. The integral over the azimuthal angle $\varphi''$ of
$\mathbf{r}''$ is trivial now. The remaining double integral over the
spherical variables $r''$ and $\theta''$ may be rewritten by introducing
$t$ and $u=r''\cos\theta''$ as new integration variables, so that $\chi$ is
determined by the relation $t\cos\chi=r-u$. The integration over $u$ can be
carried out straightforwardly. After insertion of the components
$G_{\mathrm{s},\parallel}$ and $G_{\mathrm{s},\perp}$ of the scattering
Green function \cite{SvW11}, the $zz$-component of (\ref{12}) leads to an
expression for the second moment of the $z$-component of the electric field
of the same form as in (\ref{15}), with $F_{\mathrm{c},\parallel}(r)$ given
as
\begin{eqnarray}
&&F_{\mathrm{c},\parallel}(r)=-\frac{9}{8k^3a^3}\sum_{l=1}^\infty l(l+1)
\mathrm{Re}\left[(-i)^{l+1} B^e_l J^e_{ \mathrm{c},\parallel,l}(kr)\right.\nonumber\\
&&\left.+(-i)^{l+1} B^m_l J^m_{ \mathrm{c},\parallel,l}(kr)\right]\, . \label{25}
\end{eqnarray}
The functions $J^e_{ \mathrm{c},\parallel,l}$ and 
$J^m_{\mathrm{c},\parallel,l}$ are integrals over $t$:
\begin{eqnarray}
&&J^e_{ \mathrm{c},\parallel,l}(\zeta)=\int_{\zeta-\rho}^{\zeta+\rho} dt\,
\frac{1}{t}
\biggl\{
\left[g_1(t,\zeta,\rho)-g_2(t,\zeta,\rho)\right]  \bar{H}_l(t)
\biggr.
\nonumber\\
&&
\biggl.
+2l(l+1) g_2(t,\zeta,\rho)
H_l(t)
\biggr\}
\, ,\nonumber\\
\label{26}\\
&&J^m_{ \mathrm{c},\parallel,l}(\zeta)=\int_{\zeta-\rho}^{\zeta+\rho} dt\,
t \left[g_1(t,\zeta,\rho)-g_2(t,\zeta,\rho)\right]
H_l(t)
\, , \nonumber\\
\label{27}
\end{eqnarray}
with $\rho=kR$. The auxiliary functions $g_i$ are 
\begin{eqnarray}
&&g_1(t,\zeta,\rho)=\frac{\rho^2-(\zeta-t)^2}{2\zeta}\, , \label{28}\\
&&g_2(t,\zeta,\rho)=\frac{1}{3t^2}\left(\frac{\rho^2-\zeta^2-t^2}{2\zeta}\right)^3
  +\frac{1}{3} t \, , \label{29}
\end{eqnarray}
while $H_l(t)$ and $\bar{H}_l(t)$ have been defined below (\ref{17}).

Similarly, the $xx$-component of the integrand in (\ref{12}) can be written
as $\sin^2\chi\cos^2\varphi'' \, G_{\mathrm{s},\parallel}(t,\omega+i0)+
(\cos^2\chi\cos^2\varphi''+\sin^2\varphi'')G_{\mathrm{s},\perp}(t,\omega+i0)$. Taking
the same steps as above one arrives at an expression for the second moment
of the $x$-component of the electric field as in (\ref{15}). Here,
$F_{\mathrm{c},\perp}(r)$ is found as
\begin{eqnarray}
&&F_{\mathrm{c},\perp}(r)=-\frac{9}{16k^3a^3}\sum_{l=1}^\infty l(l+1)
\mathrm{Re}\left[(-i)^{l+1} B^e_l J^e_{ \mathrm{c},\perp,l}(kr)\right.\nonumber\\
&&\left.+(-i)^{l+1} B^m_l J^m_{ \mathrm{c},\perp,l}(kr)\right]\, , \label{30}
\end{eqnarray}
with the integrals:
\begin{eqnarray}
&&J^e_{ \mathrm{c},\perp,l}(\zeta)=\int_{\zeta-\rho}^{\zeta+\rho} dt\,
\frac{1}{t}
\biggl\{
\left[g_1(t,\zeta,\rho)+g_2(t,\zeta,\rho)\right] 
\biggr.
\bar{H}_l(t)\nonumber\\
&&\left.+2l(l+1) \left[g_1(t,\zeta,\rho)-g_2(t,\zeta,\rho)\right]
H_l(t)\right\}\, ,
\label{31}\\
&&J^m_{\mathrm{c},\perp,l}(\zeta)=\int_{\zeta-\rho}^{\zeta+\rho} dt\,
t \left[g_1(t,\zeta,\rho)+g_2(t,\zeta,\rho)\right]
H_l(t)\, . \nonumber\\
\label{32}
\end{eqnarray}

For large $R$ the above results are consistent with those found previously
\cite{SvW11} for the correction functions of the atomic decay rate in the
vicinity of a halfspace.  In fact, defining $\zeta'=\zeta-\rho$ as the
(scaled) distance from the atom to the surface of the spherical domain, one
finds the asymptotic forms:
\begin{equation}
\lim_{\rho\rightarrow\infty}g_1(t,\zeta,\rho) = t-\zeta' \quad , \quad 
\lim_{\rho\rightarrow\infty} g_2(t,\zeta,\rho) =\frac{1}{3}t-\frac{\zeta'^3}{3t^2}\, . \label{33}
\end{equation}
Substituting these expressions into (\ref{26})--(\ref{27}) and
(\ref{31})--(\ref{32}) one obtains decay-rate correction functions that are
in agreement with the results of \cite{SvW11}.

For hot scatterers similar methods may be used to derive integral
representations for $F_{\mathrm{d},\parallel}(r)$ and
$F_{\mathrm{d},\perp}(r)$. To that end one starts from (\ref{20}) and
inserts the spherical-coordinate representations of the vector spherical
wave functions in terms of the spherical variables $t,\chi,\varphi''$
defined above. Subsequently, one rewrites the triple integral over
$\mathbf{r}''$ in terms of the integration variables $t, u, \varphi''$.
Upon carrying out the integrals over the latter two variables one gets
\begin{eqnarray}
&&F_{\mathrm{d},\parallel}(r)=\frac{9}{8k^3a^3}\sum_{l=1}^\infty
l(l+1)\left[C^e_l J^e_{\mathrm{d},\parallel,l}(kr)\right.\nonumber\\
&&\left. +C^m_l J^m_{\mathrm{d},\parallel,l}(kr)\right]\, , \label{34}\\
&&F_{\mathrm{d},\perp}(r)=\frac{9}{16k^3a^3}\sum_{l=1}^\infty
l(l+1)\left[C^e_l J^e_{\mathrm{d},\perp,l}(kr)\right.\nonumber\\
&&\left. +C^m_l J^m_{\mathrm{d},\perp,l}(kr)\right]\, , \label{35}
\end{eqnarray}
with integrals $J^q_{\mathrm{d},p,l}$ (for $p=\, \parallel,\perp$ and $q=e,m$)
that follow from (\ref{26})-(\ref{27}) and (\ref{31})-(\ref{32}) by
replacing $H_l(t)$ and $\bar{H}_l(t)$ with $H_{\mathrm{d},l}(t)$ and
$\bar{H}_{\mathrm{d},l}(t)$, defined below (\ref{21}).
Hence, the expressions for the hot-scatterer integrals are quite similar to
those for cold scatterers, with squares of moduli of spherical Hankel
functions (and their derivatives) occurring instead of ordinary squares, as
in the previous section.

The integral representations as given above are a suitable starting-point
to derive the asymptotic behavior of the decay-rate correction functions
for $r$ tending to $\infty$. When $r$ increases, the variable $\zeta=kr$ in
the $J_l$-integrals (\ref{26})-(\ref{27}) and (\ref{31})-(\ref{32}) gets
large. Hence, the integration variable $t$ in these integrals is large as
well, so that one may insert the asymptotic form \cite{NIST10} of the
spherical Hankel function $h^{(1)}_l(t)\simeq (-i)^{l+1} e^{it}/t$ and its derivative
$d[th^{(1)}_l(t)]/dt\simeq (-i)^l e^{it}$ in the integrands. Instead of $t$
we now introduce the new integration variable $x$, by writing $t=\zeta+x$,
so that the integration limits for $x$ become $\pm \rho$. For large $\zeta$
one may replace the combinations of $g_1$ and $g_2$ in the integrands by
the leading terms in their power series in $1/\zeta$, at fixed values of
$\rho$ and $x$. For instance, to determine the asymptotic form of (\ref{26})
we write
\begin{eqnarray}
&&\frac{1}{t}[g_1(t,\zeta,\rho)-g_2(t,\zeta,\rho)]\simeq (\rho^2-x^2)^2/(4\zeta^4)
\, , \label{36}\\
&& \frac{1}{t^3} g_2(t,\zeta,\rho)\simeq
(\rho^2-x^2)/(2\zeta^4)\, . \label{37}
\end{eqnarray}
Upon evaluating the (trivial) integral over $x$ we find:
\begin{eqnarray}
&&J^e_{ \mathrm{c},\parallel,l}(\zeta)\simeq (-1)^l
  \frac{e^{2i\zeta}}{\zeta^4}\left\{\left(-\frac{1}{2}\rho^2+\frac{3}{8}\right)\sin(2\rho)
-\frac{3}{4}\rho\cos(2\rho)\right.\nonumber\\
&&\left.-2l(l+1)\left[\frac{1}{4}\sin(2\rho)-\frac{1}{2}\rho\cos(2\rho)\right]\right\}\,
     \label{38}
\end{eqnarray}
for large $\zeta$.  In a similar way we get the asymptotic forms of the
remaining cold-scatterer integrals for large $\zeta$:
\begin{eqnarray}
J^m_{ \mathrm{c},\parallel,l}(\zeta)&&\simeq (-1)^{l+1}
  \frac{e^{2i\zeta}}{\zeta^4}
\left[\left(-\frac{1}{2}\rho^2+\frac{3}{8}\right)\sin(2\rho)\right.\nonumber\\
&&\left.-\frac{3}{4}\rho\cos(2\rho)\right]\,
, \label{39}\\
J^e_{ \mathrm{c},\perp,l}(\zeta)&&\simeq -J^m_{ \mathrm{c},\perp,l}(\zeta)\simeq
\nonumber\\
&&\simeq (-1)^l
 \frac{e^{2i\zeta}}{\zeta^2}
\left[\frac{1}{2}\sin(2\rho)-\rho\cos(2\rho)\right]\, . \label{40}
\end{eqnarray}

The asymptotic decay-rate correction functions $F_{\mathrm{c},p}(r)$ (with
$p=\, \parallel,\perp$) for cold scatterers, which follow by inserting the
above asymptotic forms of the $J_l$-integrals in (\ref{25}) and (\ref{30}),
are damped and oscillating, with a period of (half of) the wavelength. The
damping is algebraic; it is proportional to $1/r^4$ for the longitudinal
polarization, but slower, namely proportional to $1/r^2$, for the
transverse polarization. For the latter polarization all contributions of
the electric and magnetic multipoles are modulated by the same function of
$\rho$ (or the radius $R$ of the spherical domain). Hence, when $\rho$ is a
solution of the transcendental equation $\tan(2\rho)=2\rho$, the asymptotic
form of $F_{\mathrm{c},\perp}(r)$ vanishes, which means that the decay is
faster than $1/r^2$ in that case. Indeed, by evaluating the next order in
the asymptotic expansion one finds a decay proportional to $1/r^3$ for
these special values of $R$.  Hence, the scatterers are less effective in
modifying the atomic decay rate for these particular radii of the
spherical domain.

For hot scatterers the asymptotic forms of the $J_l$-integrals can be
determined in a similar fashion. Since only the absolute value of the
spherical Hankel functions enter the integrands, no oscillations are
found. One gets the following asymptotic results for large $\zeta$:
\begin{eqnarray}
J^e_{\mathrm{d},\parallel,l}(\zeta)&&\simeq\frac{1}{\zeta^4} 
\left(\frac{4}{15}\rho^5+\frac{4}{3}l(l+1)\rho^3\right)\,
, \label{41}\\
J^m_{\mathrm{d},\parallel,l}(\zeta)&&\simeq \frac{4}{15}
\frac{\rho^5}{\zeta^4}\, , \label{42}\\
 J^e_{\mathrm{d},\perp,l}(\zeta)&&\simeq J^m_{\mathrm{d},\perp,l}(\zeta)\simeq
 \frac{4}{3}\frac{\rho^3}{\zeta^2}\, . \label{43}
\end{eqnarray}
The ensuing asymptotic forms of the dielectric decay-rate correction
functions show a monotonously damped behavior proportional to $1/r^4$
and $1/r^2$ for the longitudinal and transverse polarizations,
respectively.

For positions near the surface of the spherical domain containing the
scatterers the decay-rate correction functions turn out to diverge. For
cold scatterers and a longitudinal atomic polarization this follows by
considering the integrals $J^q_{\mathrm{c}, \parallel,l}(\zeta)$ (with
$q=e,m$) for large $l$. Since for these values the spherical Hankel
function gets the form $h^{(1)}_l(t)\simeq -i\, 2^{l+1/2}\,l^l/(e^l\,
t^{l+1})$ \cite{NIST10}, the integrals (\ref{26}) and (\ref{27}) can be
evaluated explicitly. One gets:
\begin{equation}
J^e_{\mathrm{c}, \parallel,l}(\zeta)\simeq -\frac{2^{2l}\, l^{2l}}{e^{2l}}\,
    \left[ \frac{\rho}{\zeta(\zeta-\rho)^{2l+1}}-(\rho\rightarrow
      -\rho)\right] \label{44}
\end{equation}
and a similar form for $J^m_{ \parallel,l}(\zeta)$, which is found to be
proportional to $l^{2l-3}$ instead of $l^{2l}$.  Hence, for increasing
values of $l$ both integrals get large. However, in the correction function
(\ref{25}) these integrals are multiplied by the multipole amplitudes and
summed over all $l$. For large $l$ the electric-multipole amplitude (as
given by (\ref{A2})) gets the asymptotic form
\begin{equation}
\mathrm{Re}\, \left[(-i)^{l+1} B^e_l\right]\simeq \frac{\mathrm{Im}\,
  \varepsilon(\omega+i0)}{|1+\varepsilon(\omega+i0)|^2}\frac{e^{2l}\, (ka)^{2l+1}}{2^{2l}\,
  l^{2l+2}}\, ,\label{45}
\end{equation}
whereas the asymptotic form of the magnetic-multipole amplitude is
proportional to $1/l^{2l+4}$.  Upon inserting (\ref{44}) and (\ref{45})
into (\ref{25}) and summing over $l$ one finds that the contributions of
the electric multipoles diverge for $\zeta\rightarrow \rho-ka$, as it
becomes a geometric series in $ka/(\zeta-\rho)$. In contrast, the
contributions of the magnetic multipoles stay finite. As a result we obtain
the following asymptotic form of the correction function
$F_{\mathrm{c},\parallel}(r)$ for $r\rightarrow R+a$:
\begin{equation}
F_{\mathrm{c},\parallel}(r)\simeq \frac{9}{16k^3a^2}\frac{\mathrm{Im}\,
  \varepsilon(\omega+i0)}{|1+\varepsilon(\omega+i0)|^2}\frac{R}{(R+a)(r-R-a)}
\, ,\label{46}
\end{equation}
so that it diverges linearly near the surface of the spherical
domain. Since the scattering spheres may protrude from the domain, the
actual value of $r$ at which the correction function diverges is not $R$,
but $R+a$.

The correction function for the transverse atomic polarization has a
similar divergency, with a prefactor that is smaller by a factor of two:
$F_{\mathrm{c},\perp}(r)\simeq \half F_{\mathrm{c},\parallel}(r)$. These
asymptotic forms of the decay-rate correction functions for an atom near
the surface of a spherical domain filled with cold scatterers are in
agreement with those found for an atom near a halfspace with such
scatterers \cite{SvW11}, as could have been expected.

Finally, we turn to the asymptotic expressions for the decay-rate
correction functions pertaining to hot scatterers. Near the surface of the
spherical domain containing the scatterers these are found to have the same
divergency as those for cold scatterers: $F_{\mathrm{d},\parallel}(r)
\simeq F_{\mathrm{c},\parallel}(r)$ and $F_{\mathrm{d},\perp}(r) \simeq
F_{\mathrm{c},\perp}(r)$, for $r\rightarrow R+a$. The subdominant terms in
the asymptotic expressions, which are proportional to $\log(r-R-a)$, are
found to agree as well. As a consequence, the `radiative' decay-rate
correction functions $F_{\mathrm{r},\parallel}(r)$ and
$F_{\mathrm{r},\perp}(r)$ stay finite near the surface of the domain.

The linearly divergent behavior of the decay-rate correction functions
$F_{\mathrm{c},p}$ and $F_{\mathrm{d},p}$ (with $p=\parallel,\perp$) near
the surface of a spherical domain filled with scatterers is less pronounced
than that found for the atomic decay rate near a single homogeneous sphere
\cite{DKW01}. In fact, the latter is proportional to $1/(r-a)^3$ for a
sphere of radius $a$. The collective effect of an aggregate of scattering
spheres is mitigated as a consequence of the configurational averaging,
which implies a reduced probability for an individual scatterer to be near
the atom even when $r_a$ is close to $R+a$.

\section{Evaluation of the decay-rate correction 
functions\label{sectionevaluation}}

The decay-rate correction functions $F_{\mathrm{c},\parallel}$
and$F_{\mathrm{c},\perp}$ for cold scatterers, as given by (\ref{25}) and
(\ref{30}), depend on the distance $r$ via the integrals
$J^q_{\mathrm{c},p,l}$, with $p=\, \parallel,\perp$ and $q=e,m$. Inspection
of (\ref{26}), (\ref{27}), (\ref{31}) and (\ref{32}) shows that these
integrals can be written as linear combinations of a set of basic integrals
over the product of two spherical Hankel functions with an algebraic
prefactor:
\begin{equation}
I_{l_1,l_2,n}(z)=\int^z du\, u^{-n} h^{(1)}_{l_1}(u)\,
h^{(1)}_{l_2}(u)\, ,
\label{47}
\end{equation}
for $z>0$. As an example, one may write $J^m_{\mathrm{c},\parallel,l}$ as a
linear combination of $I_{l,l,n}(z)$ with $n=-5,-3,-2,-1,1$ and
$z=\zeta\pm\rho$.  Clearly, only `diagonal' basic integrals, with
$l_1=l_2$, show up here. However, the arguments of the integrals
$J^e_{\mathrm{c},\parallel,l}$ and $J^e_{\mathrm{c},\perp,l}$ in (\ref{26})
and (\ref{31}) contain both the square of a spherical Hankel function
$h^{(1)}_l(t)$ and of the derivative $d[th^{(1)}_l(t)]/dt$. The latter may
be rewritten in terms of a spherical Hankel function with a different order
\cite{NIST10}, since one has
$d[th^{(1)}_l(t)]/dt=(l+1)h^{(1)}_l(t)-th^{(1)}_{l+1}(t)$. As a
consequence, upon rewriting $J^e_{\mathrm{c},\parallel,l}$ and
$J^e_{\mathrm{c},\perp,l}$ in terms of basic integrals one encounters
off-diagonal basic integrals with $l_2=l_1+1$ as well. It turns out that
one needs information about the diagonal basic integrals $I_{l,l,n}(z)$
with $n=-5, -3, -2, -1, 0, 1, 3$ and the off-diagonal integrals
$I_{l,l+1,n}(z)$ with $n=-4, -2, -1, 0, 2$. All information about these
basic integrals is collected in Appendix \ref{appendixb}.

For hot scatterers, the integrals $J^q_{\mathrm{d},p,l}$ (with
$p=\, \parallel,\perp$ and $q=e,m$) may be analyzed in a similar way. They can
be written as linear combinations of a second set of basic integrals
containing the product of a spherical Hankel function and its complex conjugate:
\begin{equation}
I'_{l_1,l_2,n}(z)=\int^z du\, u^{-n} h^{(1)}_{l_1}(u)\,
[h^{(1)}_{l_2}(u)]^\ast \, .
\label{48}
\end{equation}

Upon collecting all information about the integrals (\ref{47}) and
(\ref{48}) we can derive explicit expressions for $J^q_{\mathrm{c},p,l}$
and $J^q_{\mathrm{d},p,l}$, for $p=\, \parallel,\perp$ and $q=e,m$. Combining
these with the expressions for the electric and magnetic multipole
amplitudes (\ref{A2}) we are able to draw curves for the decay-rate
correction functions. 

\begin{figure}[h]
  \begin{center}
   \includegraphics[height=5.5cm]{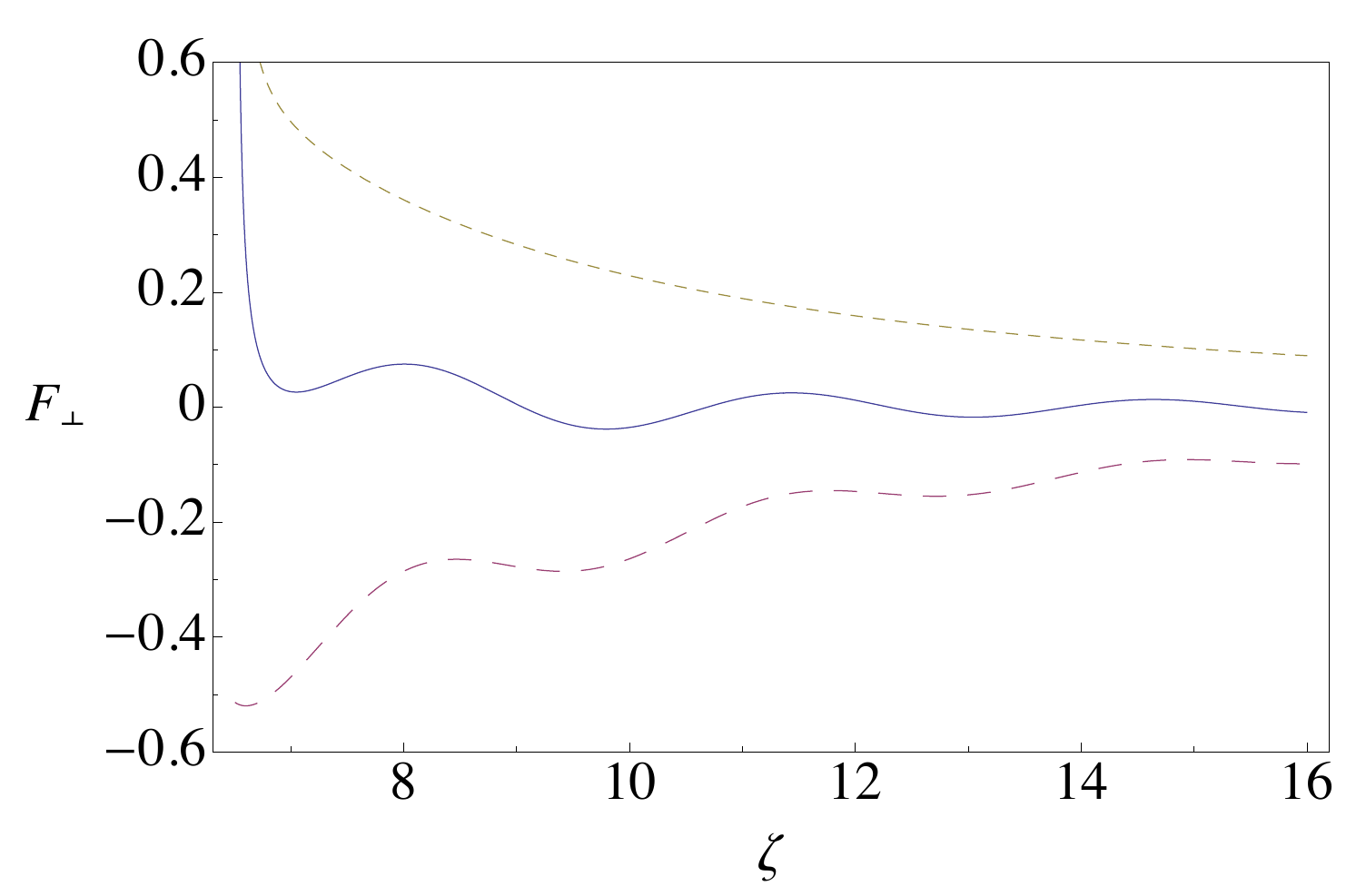}
   \caption{(Color online) Decay-rate correction function $F_{\perp}(\zeta)$
     (with $\zeta=kr$) for cold scatterers and its constituents
     $F_{\mathrm{d},\perp}(\zeta)$ ($\shortdash\,\, 
     \shortdash\,\, \shortdash\,$) and $F_{\mathrm{r},\perp}(\zeta)$
     ($\longdash\,\, \longdash\,\, \longdash\,$), for a spherical domain with
     dimensionless radius $kR=6$ containing cold absorbing spheres of
     dimensionless radius $ka=0.5$ and dielectric constant
     $\varepsilon=3.0+0.5 i$.}
\label{fig1}
  \end{center}
\end{figure}

For cold scatterers and a transverse atomic polarization the decay-rate
correction function $F_{\perp}=F_{\mathrm{c},\perp}$ is shown for a
representative choice of $R$, $a$ and $\varepsilon$ in Fig.~\ref{fig1}. As
remarked above (\ref{23}), the correction function $F_{\mathrm{c},\perp}$
can be split into a dielectric contribution $F_{\mathrm{d},\perp}$ and a
radiative contribution $F_{\mathrm{r},\perp}$.
Both $F_{\mathrm{c},\perp}$ and its constituent $F_{\mathrm{r},\perp}$ show
a characteristic damped oscillating behavior, which is proportional to
$\zeta^{-2}$ for large values of $\zeta$, in accordance with the asymptotic
form given by (\ref{30}) with (\ref{40}). In contrast, the dielectric
contribution $F_{\mathrm{d},\perp}$ is a monotonous function, which falls
off like $\zeta^{-2}$ for large $\zeta$, as has been shown in (\ref{43}).
For atomic positions near the surface of the domain containing the
scatterers the correction functions $F_{\mathrm{c},\perp}$ and
$F_{\mathrm{d},\perp}$ diverge as $1/(\zeta-kR-ka)$, in accordance with
(\ref{46}), whereas $F_{\mathrm{r},\perp}$ stays finite.

\begin{figure}[h]
  \begin{center}
   \includegraphics[height=5.5cm]{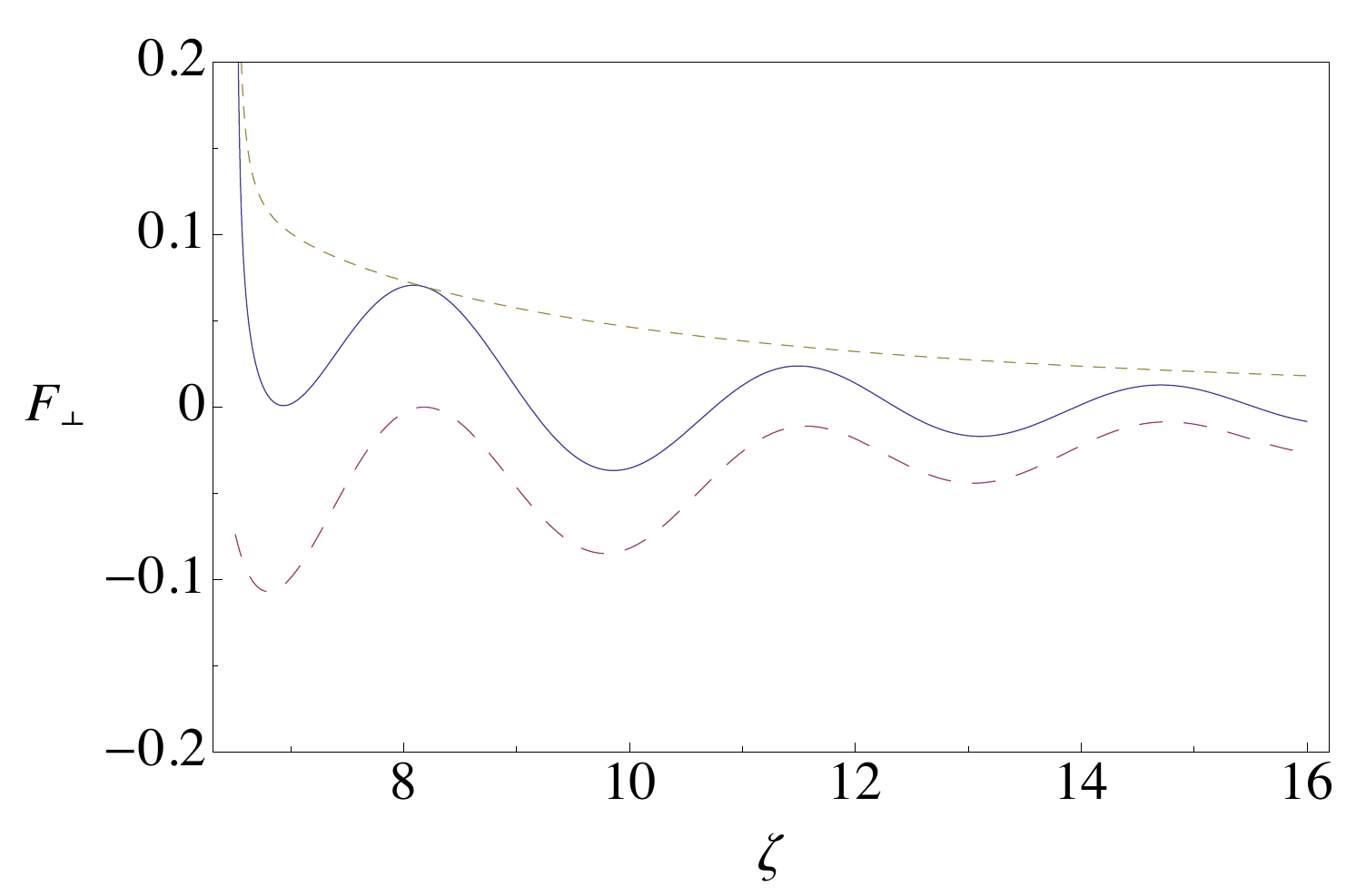}
   \caption{(Color online) Decay-rate correction function $F_{\perp}(\zeta)$
     for cold scatterers and its constituents $F_{\mathrm{d},\perp}(\zeta)$
     ($\shortdash\,\, \shortdash\,\, \shortdash\,$) and
     $F_{\mathrm{r},\perp}(\zeta)$ ($\longdash\,\, \longdash\,\, \longdash\,$),
     for $kR=6$, $ka=0.5$ and $\varepsilon=3.0+0.1 i$.}
\label{fig2}
  \end{center}
\end{figure}

\begin{figure}[h]
  \begin{center}
   \includegraphics[height=5.5cm]{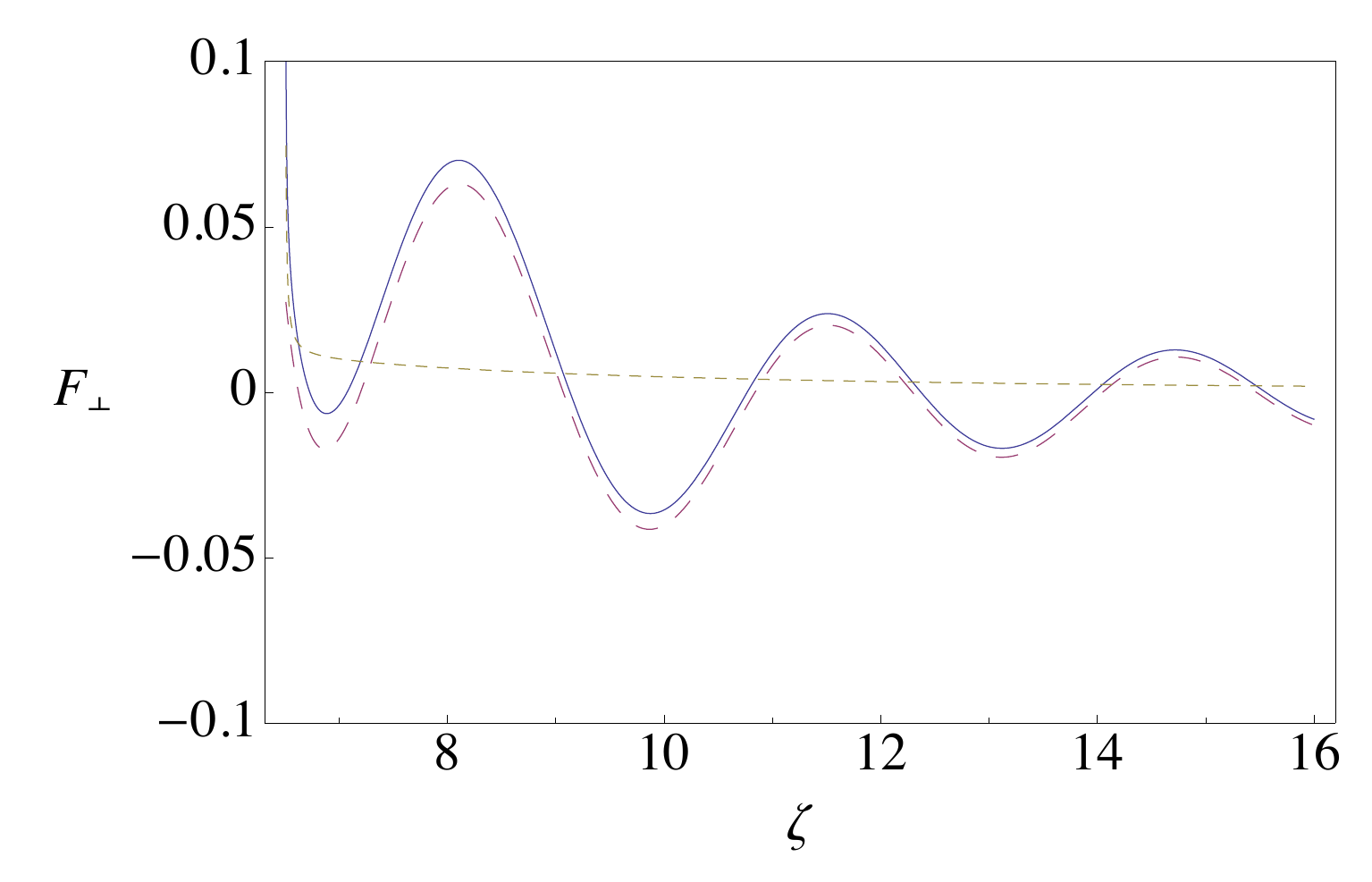}
   \caption{(Color online) Decay-rate correction function $F_{\perp}(\zeta)$
     for cold scatterers and its constituents $F_{\mathrm{d},\perp}(\zeta)$
     ($\shortdash\,\, \shortdash\,\, \shortdash\,$) and
     $F_{\mathrm{r},\perp}(\zeta)$ ($\longdash\,\, \longdash\,\, \longdash\,$),
     for $kR=6$, $ka=0.5$ and $\varepsilon=3.0+0.01 i$.}
\label{fig3}
  \end{center}
\end{figure}

The two contributions to $F_{\mathrm{c},\perp}$ are found to be mutually
counteractive: 
 $F_{\mathrm{d},\perp}$ is positive, whereas $F_{\mathrm{r},\perp}$ is
negative. Their balance shifts when the dielectric medium becomes less
absorptive, as is illustrated in Figs.~\ref{fig2},\ref{fig3}. 

\begin{figure}[h]
  \begin{center}
   \includegraphics[height=5.5cm]{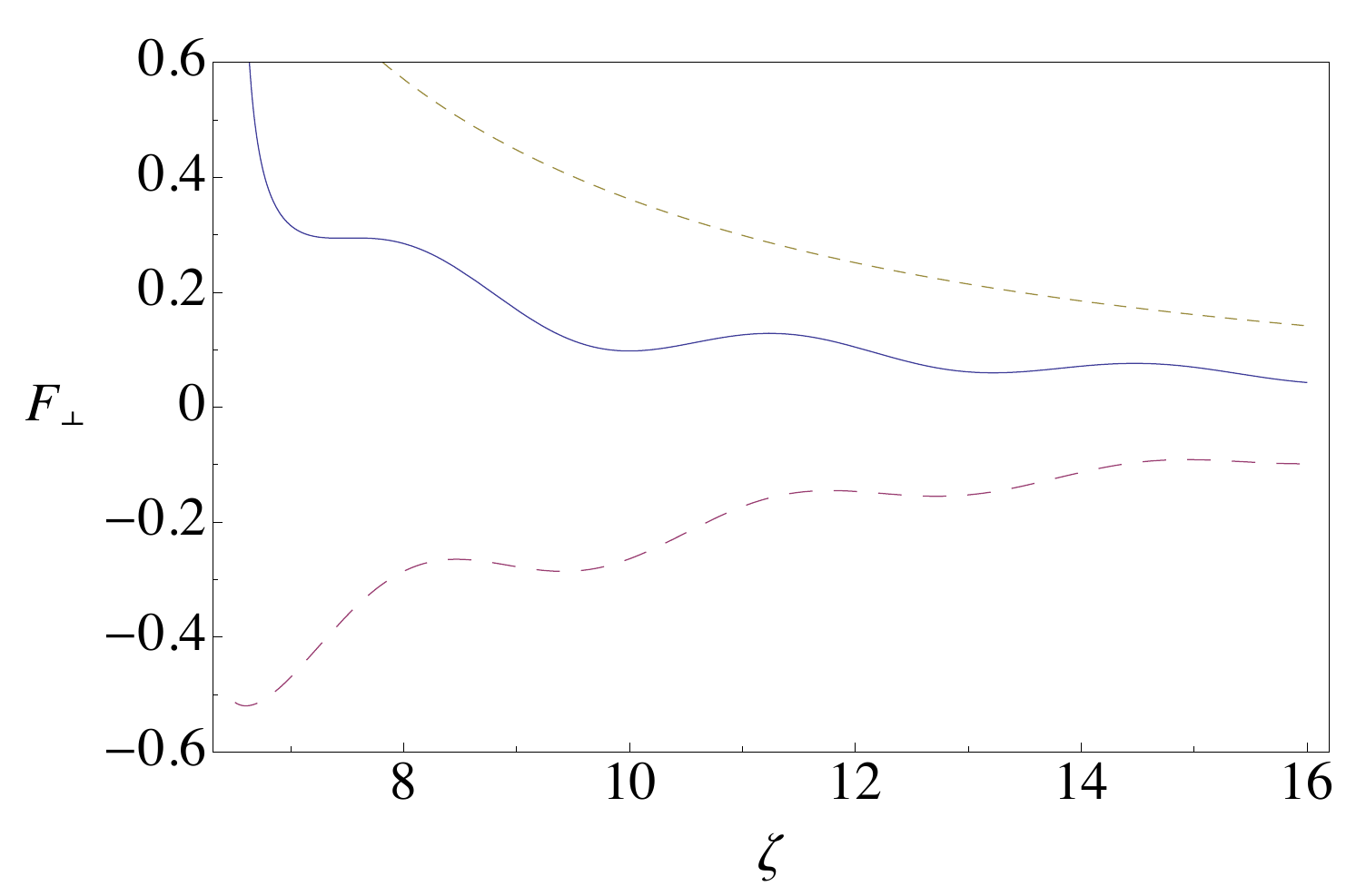}
   \caption{(Color online) Total decay-rate correction function
     $F_{\perp}(\zeta)$ for scatterers at finite temperature and its constituents
     $[e^{\beta\hbar\omega}/(e^{\beta\hbar\omega}-1)]\,
     F_{\mathrm{d},\perp}(\zeta)$ ($\shortdash\,\, \shortdash\,\, \shortdash\,$)
     and $F_{\mathrm{r},\perp}(\zeta)$
     ($\longdash\,\, \longdash\,\, \longdash\,$), for $kR=6$, $ka=0.5$,
     $\varepsilon=3.0+0.5 i$ and $\beta\hbar\omega=1$.}
\label{fig4}
  \end{center}
\end{figure}

For hot scatterers the total decay-rate correction function
$F_{\perp}$ is equal to the sum of $F_{\mathrm{r},\perp}$ and
$[e^{\beta\hbar\omega}/(e^{\beta\hbar\omega}-1)]\, F_{\mathrm{d},\perp}$,
according to (\ref{23}). In Fig.~\ref{fig4} an example of the total
correction function for hot scatterers is represented. Comparison with
Fig.~\ref{fig1} shows how the balance between the dielectric and the
radiative contributions gets shifted.

\begin{figure}[h]
  \begin{center}
   \includegraphics[height=5.5cm]{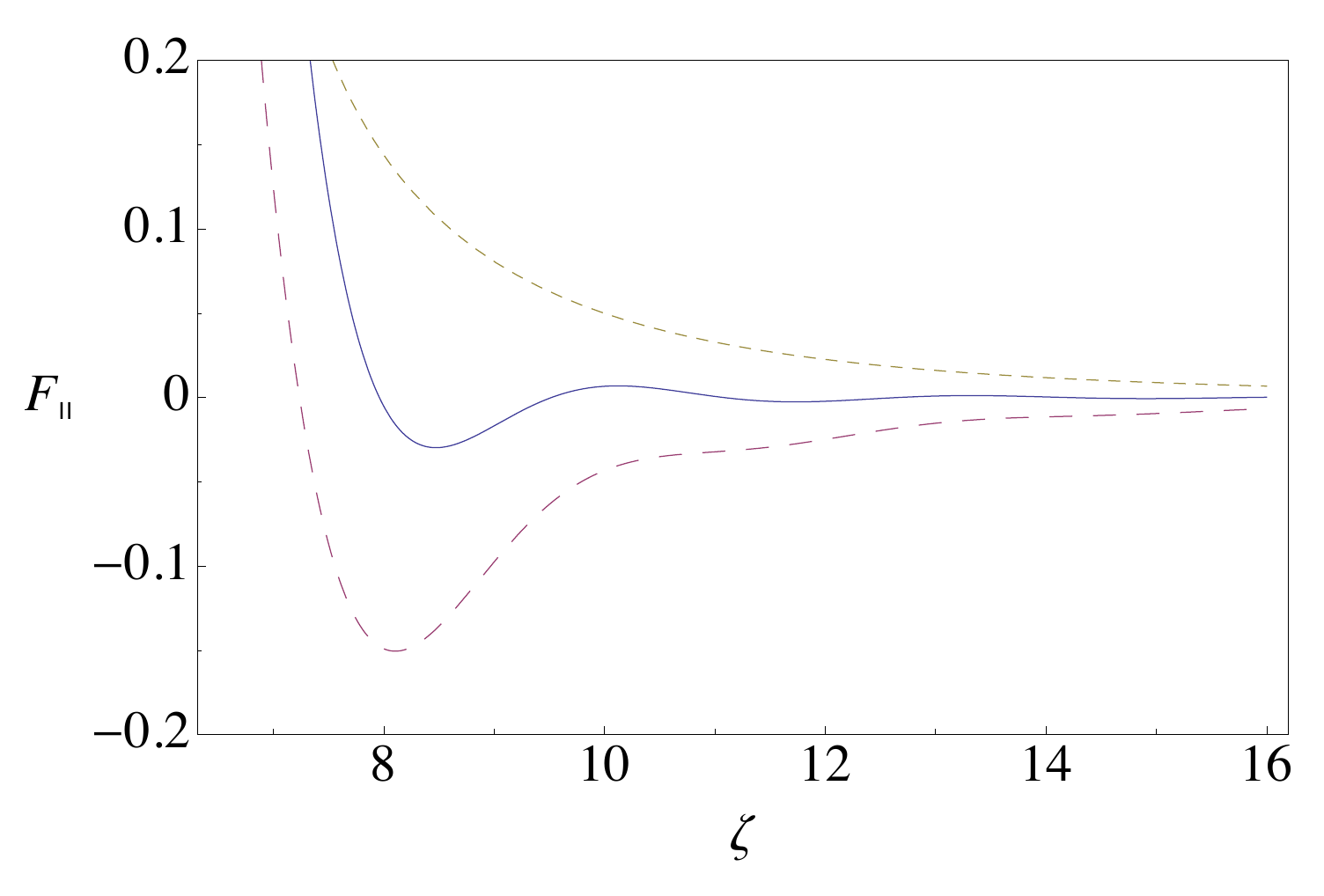}
   \caption{(Color online) Decay-rate correction function
     $F_{\parallel}(\zeta)$ for cold scatterers and its
     constituents $F_{\mathrm{d},\parallel}(\zeta)$ ($\shortdash\,\, 
     \shortdash\,\, \shortdash\,$) and $F_{\mathrm{r},\parallel}(\zeta)$
     ($\longdash\,\, \longdash\,\, \longdash\,$), for $kR=6$, $ka=0.5$ and
     $\varepsilon=3.0+0.5 i$.}
\label{fig5}
  \end{center}
\end{figure}

\begin{figure}[h]
  \begin{center}
   \includegraphics[height=5.5cm]{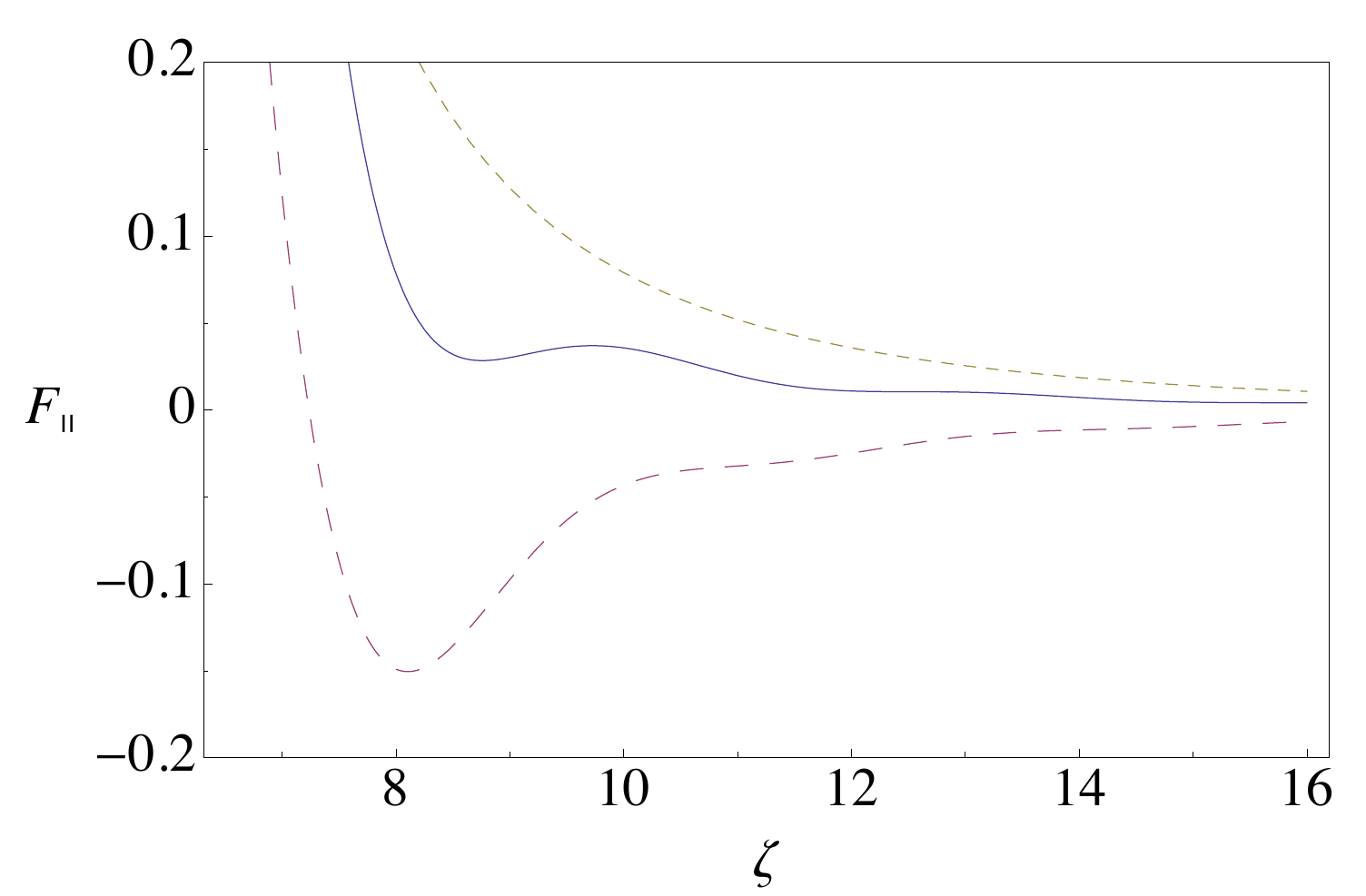}
   \caption{(Color online) Total decay-rate correction function $F_{\parallel}(\zeta)$ for
     scatterers at finite temperature and its constituents
     $[e^{\beta\hbar\omega}/(e^{\beta\hbar\omega}-1)]\,
     F_{\mathrm{d},\parallel}(\zeta)$ ($\shortdash\,\,
     \shortdash\,\,\shortdash\,$) and $F_{\mathrm{r},\parallel}(\zeta)$
     ($\longdash\,\, \longdash\,\, \longdash\,$), for $kR=6$, $ka=0.5$,
     $\varepsilon=3.0+0.5 i$ and $\beta\hbar\omega=1$.}
\label{fig6}
  \end{center}
\end{figure}

The decay-rate correction function $F_{\parallel}$ for the longitudinal
polarization direction largely behaves in a similar way, as can be seen in
Figs. \ref{fig5}--\ref{fig6}. However, in this case the oscillatory
behavior for large distances shows a faster decay, with a damping
proportional to $\zeta^{-4}$ in accordance with (\ref{25}) and
(\ref{38})--(\ref{39}). The dielectric contribution
$F_{\mathrm{d},\parallel}$ decays monotonously like $\zeta^{-4}$, as given
in (\ref{41})--(\ref{42}). The counterbalancing of the dielectric and
radiative contributions occurs here as well, with a shifting balance as the
absorptive power of the scatterers gets weaker. The inversely-linear
divergent behavior for small distances, as given by (\ref{46}), is clearly
visible. In Fig.\ \ref{fig6} it is shown how thermal radiation from the
scatterers influences the longitudinal decay-rate correction function: both
the total correction function and its dielectric part become larger owing
to stimulated-emission effects, whereas the radiative part remains
unchanged.

The present results for a spherical domain of radius $R$ filled with a
dilute set of scatterers may be compared to those obtained from an
effective-medium theory in which the domain contains a uniform dielectric
with an effective dielectric constant $\varepsilon^{(\mathrm{eff})}$. For
scatterers with a size much smaller than the wavelength one may use
Maxwell Garnett theory \cite{MG04}, with the effective dielectric constant
$\varepsilon^{(\mathrm{eff})}=1+3f(\varepsilon-1)/(\varepsilon+2)$. For a
set of spherical scatterers with a radius $a$ that is not small compared to
the wavelength (as in our model) the effective dielectric constant will
depend on $ka$, as has been discussed in \cite{D89,R00}.  It may be chosen
as $\varepsilon^{(\mathrm{eff})}=1-3ifB^e_1/(k^3a^3)$, with $B^e_1$ the
electric-dipole scattering amplitude for a sphere with radius $a$ and
dielectric constant $\varepsilon$, as given in (\ref{A2}). In the limit of
small $ka$ the latter expression reduces to the Maxwell Garnett form.

The decay-rate correction functions that follow from the effective-medium
theory will depend on the electric and magnetic multipole amplitudes
$B^p_l(\varepsilon^{(\mathrm{eff})},R)$ (with $p=e,m$) for a uniform sphere
with radius $R$ and dielectric constant
$\varepsilon^{(\mathrm{eff})}$. Since the latter differs from 1 by a small
amount, we may write the amplitudes $B^p_l(\varepsilon^{(\mathrm{eff})},R)$
as $(\varepsilon^{(\mathrm{eff})}-1)\bar{B}^p_l$, with reduced amplitudes
$\bar{B}^p_l$ depending on $\rho=kR$:
\begin{eqnarray}
 && \bar{B}^e_l=i^l\frac{2l+1}{l(l+1)}\rho\left\{(l+1+\half\rho^2)[j_l(\rho)]^2+ 
\half\rho^2[j_{l+1}(\rho)]^2\right.\nonumber\\
&&\left.-(l+\threehalf)\rho j_l(\rho)j_{l+1}(\rho)\right\}\, , \label{49}\\
&&\bar{B}^m_l=i^l\frac{2l+1}{l(l+1)}\rho\left\{\half\rho^2[j_l(\rho)]^2+ 
\half\rho^2[j_{l+1}(\rho)]^2\right.\nonumber\\
&&\left.-(l+\half)\rho j_l(\rho)j_{l+1}(\rho)\right\}\, . \label{50}
\end{eqnarray}
In terms of these reduced amplitudes the effective-medium decay-rate
correction functions for cold scatterers become
\begin{eqnarray}
&&F^{(\mathrm{eff})}_{\mathrm{c},\parallel}(r)=\frac{9}{2k^3a^3}\sum_{l=1}^\infty
[l(l+1)]^2 \mathrm{Re}\left[B^e_1(-i)^l \right.\nonumber\\
&&\left.\times\bar{B}^e_l \frac{1}{k^2r^2} H_l(kr)\right]\, , \label{51}\\
&& F^{(\mathrm{eff})}_{\mathrm{c},\perp}(r)=\frac{9}{4k^3a^3}\sum_{l=1}^\infty
l(l+1)\mathrm{Re}\left\{B^e_1(-i)^l \right.\nonumber\\
&&\left.\times\left[\bar{B}^e_l \frac{1}{k^2r^2}
  \bar{H}_l(kr)+\bar{B}^m_l H_l(kr)\right]\right\}\, . \label{52}
\end{eqnarray}
The effective-medium dielectric decay-rate correction functions
$F^{(\mathrm{eff})}_{\mathrm{d},p}(r)$ (with $p=\parallel,\perp$), which
come into play for hot scatterers, have a similar form. They follow from
(\ref{51})--(\ref{52}) by replacing $H_l$ and $\bar{H}_l$ with the real
functions $H_{\mathrm{d},l}$ and $\bar{H}_{\mathrm{d},l}$ (defined below
(\ref{21})) and changing the overall sign. It should be noted that $(-i)^l
\bar{B}^q_l$ (for $q=e,m$) is real, so that the functions
$F^{(\mathrm{eff})}_{\mathrm{d},p}(r)$ are proportional to $\mathrm{Re}
(B^e_1)$.

  Having established the expressions for the effective-medium decay-rate
  correction functions, we may compare the ensuing curves with those from
  our model in which the effects of inhomogeneities were taken into
  account. As an example, we shall consider the correction functions for
  the choice of parameters considered in Fig.\ \ref{fig2}. 
\begin{figure}[h]
 \begin{center}
 \includegraphics[height=5.5cm]{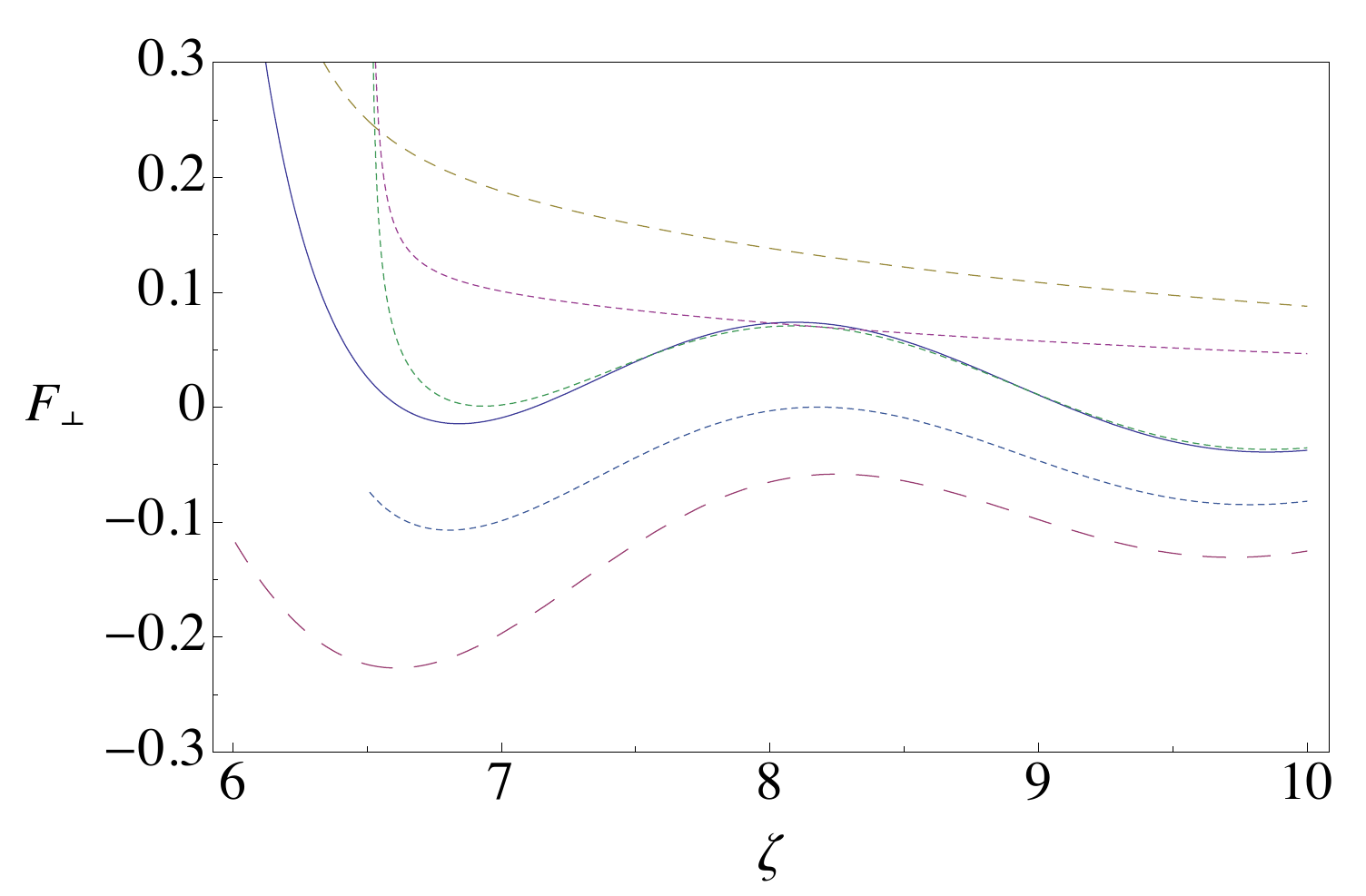}
 \caption{(Color online) Comparison of the decay-rate correction function
   $F^{(\mathrm{eff})}_{\perp}(\zeta)$
   ($\protect\rule[0.5mm]{3mm}{0.1mm}$) for a cold effective medium and its
   constituents $F^{(\mathrm{eff})}_{\mathrm{d},\perp}(\zeta)$
   ($\shortdash\,\, \shortdash\,\, \shortdash\,$),
   $F^{(\mathrm{eff})}_{\mathrm{r},\perp}(\zeta)$ ($\longdash\,\,
   \longdash\,\, \longdash\,$) with their scattering-medium counterparts
   $F_{\perp}(\zeta)$, $F_{\mathrm{d},\perp}(\zeta)$,
   $F_{\mathrm{r},\perp}(\zeta)$
   ($\points\,\points\,\points\,\points\,\points\,$) for cold scatterers,
   as taken from Fig.\ 2, for $kR=6$, $ka=0.5$ and $\varepsilon=3.0+0.1
   i$.}
\label{fig7}
\end{center}
\end{figure}
From Fig.\ \ref{fig7} it is clear that for large distances the
effective-medium theory gives adequate results for the decay-rate
correction function $F_{\perp}=F_{\mathrm{c},\perp}$ for cold scatterers,
whereas it is off (by about a factor 2) for the dielectric part
$F_{\mathrm{d},\perp}$ and also for the radiative part
$F_{\mathrm{r},\perp}$, which play a role for hot scatterers. Thus, the
effective-medium theory is not able to predict correct values for the
decay-rate correction functions in the presence of hot scatterers. This
state of affairs could have been expected from the general arguments put
forward in \cite{R00}.

In the near zone the predictions of the effective-medium theory get even
less reliable. As remarked at the end of Section \ref{sectionintegral}, the
small-distance asymptotic behavior of the decay-rate correction functions
near a domain containing a set of discrete scatterers is different from
that for a uniformly-filled domain. Moreover, the effective-medium theory
misses the contribution of all higher-multipole scattering amplitudes,
which get important near the scatterers. The resulting differences are
clearly visible in the curves in Fig.\ \ref{fig7}.

  Further insight in the cause of the discrepancies found in Fig.\
  \ref{fig7} is gained by determining the asymptotic behavior of the
  effective-medium decay-rate correction functions for large
  $kr$. Substituting the asymptotic form of the spherical Hankel functions
  contained in $H_l(kr)$ and $\bar{H}_l(kr)$ in (\ref{52}), we get an
  expression for $F^{(\mathrm{eff})}_{\mathrm{c},\perp}(r)$ that is
  proportional to $\sum_{l=1}^\infty l(l+1) i^l
  (\bar{B}^e_l-\bar{B}^m_l)$. Inserting (\ref{49})--(\ref{50}), using the
  recurrence relations for $j_l$ \cite{NIST10} and employing the sum rule
  (\ref{C10}) we get the asymptotic result
\begin{equation}
F^{(\mathrm{eff})}_{\mathrm{c},\perp}(r)\simeq -\frac{9}{8k^3a^3}\left[
  \frac{1}{2}\sin (2\rho)-\rho\cos(2\rho)\right]\mathrm{Re}\left(B^e_1
  \frac{e^{2i\zeta}}{\zeta^2}\right) \, , \label{53}
\end{equation}  
for large $\zeta=kr$.  Comparison with (\ref{30}) and (\ref{40}) shows that we
have recovered the contribution of the electric-dipole scattering
amplitude. Clearly, it dominates the long-distance behavior of
$F^{(\mathrm{eff})}_{\mathrm{c},\perp}(r)$, at least for the rather small
dimensionless radius of the scatterers chosen here ($ka=0.5$).

The asymptotic form of $F^{(\mathrm{eff})}_{\mathrm{d},\perp}(r)$ may
likewise be determined. It is found to be proportional to the sum
$\sum_{l=1}^\infty l(l+1) (-i)^l(\bar{B}^e_l+\bar{B}^m_l)$, which may be
evaluated with the help of (\ref{C7})--(\ref{C9}). One gets the asymptotic
form for large $\zeta$:
\begin{equation}
  F^{(\mathrm{eff})}_{\mathrm{d},\perp}(r)\simeq -  \frac{3}{2k^3a^3}\frac{\rho^3}{\zeta^2}
\mathrm{Re}(B^e_1) \, . \label{54}
\end{equation}  
Comparing with (\ref{35}) and (\ref{43}) we see that even the
electric-dipole contributions differ: in the effective-medium theory the
factor $C^e_1$ (defined in (\ref{A17})) is replaced by
$-\mathrm{Re}(B^e_1)$. Since these differ by a factor 2.06 in the present
case, the discrepancy in Fig.\ \ref{fig7} is fully explained. Incidentally,
it should be remarked that choosing a different form for the effective
dielectric constant (for instance that of the original Maxwell Garnett
theory) does not improve matters here. An analysis of the decay-rate
correction functions for the longitudinal polarization ($\parallel$) and
of their asymptotic behavior leads to similar conclusions.

\section{Discussion\label{sectiondiscussion}}
In the previous sections the modification of the decay rate of an excited
atom in the vicinity of a spherical aggregate of randomly distributed small
dielectric spheres at arbitrary temperature has been analyzed in detail.
The changes could be described by a set of decay-rate correction functions
that depend on three independent length scales, namely on the radii of both
the scatterers and of the domain containing them, and on the distance
between the atom and the center of the aggregate. Two types of decay-rate
correction functions have been introduced, each for two different
polarizations. The decay-rate correction functions
$F_{\mathrm{c},\parallel}$ and $F_{\mathrm{c},\perp}$ describing the
influence of cold scatterers were shown to exhibit a damped oscillatory
behavior as a function of the atomic distance. In contrast, the dielectric
correction functions $F_{\mathrm{d},\parallel}$ and $F_{\mathrm{d},\perp}$
that come into play for hot scatterers are monotonous as a function of the
distance. The latter can be used as well to split the decay-rate correction
functions for both cold and hot scatterers in contributions with a
different origin, {\em viz} a dielectric and a radiative part. As we have
seen, these two parts represent decay-modifying effects that may be
counteracting, at least for sufficiently high absorptive power of the
scatterers.

Two rather different representations for the decay-rate correction
functions have been found in Sections \ref{sectioncoldhot} and
\ref{sectionintegral}. The first of these, which has been given in
(\ref{16})-(\ref{17}), follows by using addition theorems for vector
spherical wave functions. The resulting expressions are triple sums over
terms in which the electric and magnetic multipole contributions show up on
an equal footing, while the three independent length scales are neatly
separated. In the second representation, which involves sums of integrals
over finite intervals, two of these length-scales (namely, the radius
of the domain containing the scatterers and the distance of the atom to the
center of the aggregate) get intertwined in the integrands and the boundaries of the
integrals. Furthermore, the symmetry between the electric and magnetic
contributions is no longer obvious. Nevertheless, this representation is
quite useful, as it is more suitable for an efficient numerical evaluation
of the decay-rate correction functions.

The changes in the decay rate that follow from (\ref{22}) are proportional
to the density $n_s$ of the scatterers. This is a direct consequence of the
approximations for the Green function that were introduced in (\ref{11})
and above (\ref{20}). It amounts to the neglect of multiple-scattering
events that would be important in an aggregate with a high filling
fraction. When taking into account the effects of multiple scattering one
expects a nonlinear dependence of the decay rate on the density.  As an
analytical evaluation is not feasible in that case, one has to resort to
numerical simulations in order to treat dense aggregates of spheres.

A comparison of our results with those obtained from an effective-medium
theory shows how the inhomogeneities brought about by the discrete
scatterers affect the atomic decay rate. For scatterers that are far from
the atom the dielectric correction functions $F_{\mathrm{d},p}$ and the
radiative correction functions $F_{\mathrm{r},p}$ (for $p=\perp,\parallel$)
both change considerably when the medium is smoothed, while their sums
$F_{\mathrm{c},p}$ are almost left unchanged. Hence, for cold scatterers at
large distances from the atom effective-medium theory is a useful
approximation, whereas it is unreliable for scatterers at finite
temperature. In contrast, for scatterers near the atom the predictions of
effective-medium theory are inadequate at any temperature.

As we have seen, the rate of photon absorption of a ground-state atom near
a collection of dielectric spheres emitting thermal radiation is determined
by the same dielectric decay-rate correction functions
$F_{\mathrm{d},\parallel}$ and $F_{\mathrm{d},\perp}$. If the atom is in a
mixed state involving its ground state and excited state, both atomic decay
and atomic absorption processes may occur, with rates given by (\ref{22})
and (\ref{24}). In that case a stationary situation may result with an
atomic population ratio $n_e/n_g$ determined by the equilibrium condition
$n_e\langle\Gamma_\mathrm{t}\rangle=n_g\langle
\Gamma_\mathrm{a}\rangle$. Obviously, for positions far away from the
aggregate the atom will end in its ground state, since the influence of the
scatterers gets small. For distances near the surface of the domain
containing the scatterers all correction functions become large and equal,
apart from trivial factors 2. Hence, the vacuum decay rate $\Gamma_0$ in
(\ref{22}) can be neglected, so that $\langle\Gamma\rangle$ and
$\langle\Gamma_\mathrm{a}\rangle$ are proportional in that case. As a
consequence, the population ratio $n_e/n_g$ for an atom near the surface of
the domain reduces to the standard Boltzmann factor
$e^{-\beta\hbar\omega}$. Clearly, the aggregate of scatterers imposes its
temperature on the atom in this case. For intermediate atomic positions the
population ratio at which emission and absorption processes are in
equilibrium differs from this simple result, since the influence of the
scatterers gets smaller with distance. The precise form of the population
ratio as a function of the atomic distance follows directly from the
explicit expressions for the correction functions that have been determined
in this paper.

\appendix
\section{Green functions\label{appendixa}}

The Green function in the presence of a dielectric sphere, with radius $a$
and complex dielectric constant $\varepsilon$, is the sum of the vacuum
Green function $\bfsfG_0$ and a scattering term $\bfsfG_\mathrm{s}$. For a
sphere centered at the origin the latter has the form \cite{T71,C95,LKLY94}
\begin{eqnarray}
&&\bfsfG_\mathrm{s}(\mathbf{r},\mathbf{r}',\omega+i0)=
k\sum_{l=1}^{\infty}\sum_{m=-l}^{l}
\frac{(-i)^{l}(-1)^m}{2l+1}
\nonumber\\
&&\times\left[
B^e_l\, \mathbf{N}_{l,m}^{(h)}(\mathbf{r})\mathbf{N}_{l,-m}^{(h)}(\mathbf{r}')+
B^m_l\, \mathbf{M}_{l,m}^{(h)}(\mathbf{r})\mathbf{M}_{l,-m}^{(h)}(\mathbf{r}')\right]\nonumber\\
\label{A1}
\end{eqnarray}
for $\mathbf{r}$ and $\mathbf{r}'$ both outside the sphere.  

The electric and magnetic multipole amplitudes read \cite{M08,BW99,SvW11}
\begin{equation}
B^p_l=i^{l+1}\, \frac{2l+1}{l(l+1)}\,
\frac{N^p_l}{D^p_l}\, ,
 \label{A2}
\end{equation}
with $l\geq 1$ and $p=e,m$. The numerators and denominators are given as
\begin{eqnarray}
  &&N^e_l=\varepsilon\, f_l(q)\, j_l(q')- j_l(q)\,f_l(q')\, 
  , \nonumber\\
&& N^m_l=f_l(q)\, j_l(q')-j_l(q)\,f_l(q')\, 
  , \nonumber\\ 
&&D^e_l=\varepsilon\, f^{(h)}_l(q)\, j_l(q')-h^{(1)}_l(q)\,f_l(q')\, 
  , \nonumber\\
&&D^m_l= f^{(h)}_l(q)\, j_l(q')-h^{(1)}_l(q)\,f_l(q')\,   , 
\label{A3}
\end{eqnarray}
with $f_l(q)=(l+1)\, j_l(q)-q\, j_{l+1}(q)$ and $f^{(h)}_l(q)=(l+1)\,
h^{(1)}_l(q)-q\, h^{(1)}_{l+1}(q)$. The spherical Bessel and Hankel
functions depend on $q=k a$ and $q'=\sqrt{\varepsilon}\, q$, with $a$ the
radius of the sphere.

The vector spherical wave functions are defined as
\begin{eqnarray}
&&\mathbf{M}_{l,m}(\mathbf{r})=\nabla\wedge[\mathbf{r}\psi_{l,m}(\mathbf{r})]\,
,\label{A4}\\
&&\mathbf{N}_{l,m}(\mathbf{r})=k^{-1}\nabla\wedge[\nabla\wedge[\mathbf{r}\psi_{l,m}(\mathbf{r})]]\,
,\label{A5}
\end{eqnarray}
where the scalar spherical wave function $\psi_{l,m}(\mathbf{r})$ stands
for $j_l(kr)\, Y_{l,m}(\hat{\mathbf{r}})$, with $j_l$ a spherical Bessel
function and $Y_{l,m}$ a spherical harmonic depending on the angles
$(\theta,\varphi)$ that determine the direction of the unit vector
$\hat{\mathbf{r}}=\mathbf{r}/r$. The superscripts $(h)$ in (\ref{A1})
denote the analogous vector spherical wave functions with spherical Hankel
functions $h_l^{(1)}$ instead of $j_l$.

The vector spherical wave functions $\mathbf{M}_{l,m}(\mathbf{r})$ and
$\mathbf{N}_{l,m}(\mathbf{r})$ satisfy addition theorems
\cite{FJ87,C95,S61,C62}. To derive them in a concise way \cite{HC08} one
may use an expansion for $\bfsfI\, e^{i\mathbf{k}\cdot\mathbf{r}}$ that is
a generalization of the standard Rayleigh expansion for
$e^{i\mathbf{k}\cdot\mathbf{r}}$ in terms of spherical harmonics. The
addition theorem for $\mathbf{M}_{l,m}(\mathbf{r})$ reads:
 \begin{eqnarray}
&&\mathbf{M}_{l,m}(\mathbf{r}'+\mathbf{r}'')
=4\pi\sum_{l'=1}^\infty\sum_{m'=-l'}^{l'}\sum_{l''=0}^\infty\sum_{m''=-l''}^{l''}i^{l'+l''-l}
\nonumber\\
&&\times\sqrt{\frac{l(l+1)}{l'(l'+1)}}\left[ I^{(\mathrm{e})}_{l,m,l',m',l'',m''}\mathbf{M}_{l',m'}(\mathbf{r}')
  \right.\nonumber\\
&&\left.+I^{(\mathrm{o})}_{l,m,l',m',l'',m''}
\mathbf{N}_{l',m'}(\mathbf{r}') \right]\, \psi_{l'',m''}(\mathbf{r''})\, .\label{A6}
\end{eqnarray}
The coefficients $I^{(\mathrm{p})}$ with $\mathrm{p}=\mathrm{e},\mathrm{o}$ are
\begin{eqnarray}
&&I^{(\mathrm{p})}_{l,m,l',m',l'',m''}=(-1)^{m+1}\sqrt{\frac{(2l+1)(2l'+1)(2l''+1)}{4\pi}}\nonumber\\
&&\times
\left(\begin{array}{ccc}
l & l' & l''\\
m & -m' & -m''
\end{array}\right)
\left(\begin{array}{ccc}
l & l' & l''\\
1 & -1 & 0
\end{array}\right)\, \delta^\mathrm{p}_{l+l'+l''}\, ,
\label{A7}
\end{eqnarray}
with symbols $\delta^p_l$ that have been defined below (\ref{14}).  The
coefficients (\ref{A7}) agree with those given in \cite{TPM11}, and are
equivalent with the results obtained in \cite{FJ87} and \cite{HC08}, as
follows by employing a couple of identities for 6$j$-symbols \cite{TL96}.

The addition theorem for $\mathbf{N}_{l,m}$ has the same form as
(\ref{A6}), with $\mathbf{M}$ and $\mathbf{N}$ interchanged. This follows
immediately from the identities
$\mathbf{M}_{l,m}=k^{-1}\nabla\wedge\mathbf{N}_{l,m}$ and
$\mathbf{N}_{l,m}=k^{-1}\nabla\wedge\mathbf{M}_{l,m}$. Finally, we remark
that analogous addition theorems hold true for $\mathbf{M}^{(h)}_{l,m}$ and
$\mathbf{N}^{(h)}_{l,m}$. In particular, for $r'\geq r''$ the addition
theorem for $\mathbf{M}^{(h)}_{l,m}(\mathbf{r}'+\mathbf{r}'')$ follows from
(\ref{A6}) by replacing both $\mathbf{M}$ and $\mathbf{N}$ by their
counterparts $\mathbf{M}^{(h)}$ and $\mathbf{N}^{(h)}$.

For coinciding positions $\mathbf{r}$ and $\mathbf{r}'$ the scattering term
(\ref{A1}) in the Green function is diagonal in spherical coordinates:
\begin{eqnarray}
&&\bfsfG_\mathrm{s}(\mathbf{r},
\mathbf{r},\omega+i0)=G_{\mathrm{s},\parallel}(r,\omega+i0)\mathbf{e}_r\mathbf{e}_r\nonumber\\
&&+G_{\mathrm{s},\perp}(r,\omega+i0)
(\mathbf{e}_\theta\mathbf{e}_\theta+\mathbf{e}_\varphi\mathbf{e}_\varphi)\, ,
\label{A8}
\end{eqnarray}
with components given in \cite{DKW01,SvW11}.

If $\mathbf{r}'$ is situated inside the sphere and $\mathbf{r}$ is outside,
the full Green function
$\bfsfG_0+\bfsfG_\mathrm{s}\equiv\bfsfG_{0\mathrm{s}}$ reads
\cite{T71,C95,LKLY94}
\begin{eqnarray}
&&\bfsfG_{0\mathrm{s}}(\mathbf{r},\mathbf{r}',\omega+i0)=\frac{1}{a}
\sum_{l=1}^{\infty}\sum_{m=-l}^{l}
\frac{(-i)^{l+1}(-1)^m}{2l+1}
\nonumber\\
&&\times\left[
A^e_l\, \mathbf{N}_{l,m}^{(h)}(\mathbf{r})\mathbf{N}_{l,-m}^{(\varepsilon)}(\mathbf{r}')+
A^m_l\, \mathbf{M}_{l,m}^{(h)}(\mathbf{r})
\mathbf{M}_{l,-m}^{(\varepsilon)}
(\mathbf{ r}')\right]\, .\nonumber\\
\label{A9}
\end{eqnarray}
The modified vector spherical wave functions
$\mathbf{M}_{l,m}^{(\varepsilon)}$ and $\mathbf{N}_{l,m}^{(\varepsilon)}$
follow from (\ref{A4}) and (\ref{A5}) upon replacing $k$ by
$k'=\sqrt{\varepsilon} k$. Furthermore, the amplitudes $A_l^p$ are defined
as
\begin{equation}
A_l^p=i^{l+1}\, \frac{2l+1}{l(l+1)}\,
\frac{c_p}{D^p_l}\, ,
 \label{A10}
\end{equation}
with $l\geq 1$ and $p=e,m$. The coefficients in the numerator are
$c_e=\sqrt\epsilon$, $c_m=1$, while the denominators follow from
(\ref{A3}).

In the main text we need expressions for integrals over scalar products of
the modified vector spherical wave functions. These can be evaluated by
starting from (\ref{A4})-(\ref{A5}), employing the standard expressions for
the curl of a vector in spherical coordinates and evaluating the angular
integrals. In this way one gets:
\begin{eqnarray}
&&\int_{r<a}d\mathbf{r}\,  \mathbf{M}_{l,m}^{(\varepsilon)}(\mathbf{r})\cdot
\mathbf{M}_{l',m'}^{(\varepsilon)\ast}(\mathbf{r})=\delta_{l,l'}\delta_{m,m'}
l(l+1) I_l^{(\varepsilon)}(a)\, , \nonumber\\
\label{A11}\\
&&\int_{r<a}d\mathbf{r}\,  \mathbf{N}_{l,m}^{(\varepsilon)}(\mathbf{r})\cdot
\mathbf{N}_{l',m'}^{(\varepsilon)\ast}(\mathbf{r})=\delta_{l,l'}\delta_{m,m'}
\frac{l(l+1)}{2l+1}\nonumber\\
&&\times \left[(l+1)I_{l-1}^{(\varepsilon)}(a)+
l I_{l+1}^{(\varepsilon)}(a)\right]\, , \label{A12}\\
&& \int_{r<a}d\mathbf{r}\,  \mathbf{M}_{l,m}^{(\varepsilon)}(\mathbf{r})\cdot
\mathbf{N}_{l',m'}^{(\varepsilon)\ast}(\mathbf{r})=0\, ,
\label{A13}
\end{eqnarray}
with the radial integrals $I_l^{(\varepsilon)}(a)=\int_0^a dr \,
r^2|j_l(k'r)|^2$. The latter are given as:
\begin{equation}
I_l^{(\varepsilon)}(a)=2a^2\mathrm{Re}\left[\frac{k'}{k'^2-(k'^\ast)^2}j_{l+1}(k'a)j_l(k'^\ast
  a)\right]\, ,
\label{A14}
\end{equation}
as may be verified by differentiation. This identity is useful in proving
relations between $A_l^p$ and $B^p_l$. In fact, by using the Wronskian for
spherical Bessel functions \cite{NIST10} one may derive the following
equalities for $l\geq 1$:
\begin{eqnarray}
&&\left[\mathrm{Im}\, \varepsilon(\omega+i0)\right]\, 
I_l^{(\varepsilon)}(a)
|A^m_l|^2=\frac{2l+1}{l(l+1)}\frac{a^2}{k}C_l^m\, ,
\label{A15}\\
&&\left[\mathrm{Im}\, \varepsilon(\omega+i0)\right]
\left[ (l+1) I_{l-1}^{(\varepsilon)}(a) +l
  I_{l+1}^{(\varepsilon)}(a)\right]|A^e_l|^2=\nonumber\\
&&=\frac{(2l+1)^2}{l(l+1)}\frac{a^2}{k} C_l^e\, .
\label{A16}
\end{eqnarray} 
Here we introduced the abbreviations 
\begin{equation}
C_l^p=\mathrm{Re}\left[(-i)^{l+1} B^p_l\right]-\frac{l(l+1)}{2l+1}|B^p_l|^2\,
,
\label{A17}
\end{equation}
for $p=e,m$ and $l\geq 1$.

\section{Evaluation of integrals\label{appendixb}}

In the main text we introduced the indefinite integrals
\begin{equation}
I_{l_1,l_2,n}(z)=\int^z du\, u^{-n} h^{(1)}_{l_1}(u)\,
h^{(1)}_{l_2}(u)\, ,
\label{B1}
\end{equation}
for real $z>0$, integer $n$ and non-negative integers $l_1,l_2$. We need
information about these integrals for the diagonal case with $l_1=l_2=l$
and for the off-diagonal case with $l_1=l_2-1=l$. In the following we shall
omit any additive constants which may show up when deriving expressions for
these integrals.

The diagonal integrals with $l_1=l_2=l$ satisfy a recursion relation of the
form
\begin{eqnarray}
&&(2l+n+3)I_{l+1,l+1,n}(z)=(2l-n+1) I_{l,l,n}(z)\nonumber\\
&&-z^{-n+1}\left\{\left[h^{(1)}_l(z)\right]^2+\left[h^{(1)}_{l+1}(z)\right]^2\right\}\,
,
\label{B2}
\end{eqnarray}
as may be verified by differentiation. When these diagonal integrals are
known, the non-diagonal integrals follow from the relation
\begin{equation}
I_{l,l+1,n}(z)=\half(2l-n) I_{l,l,n+1}(z)-\half
z^{-n}\left[h^{(1)}_l(z)\right]^2 , 
\label{B3}
\end{equation}
which likewise may be established by differentiation. 

The recursion relation (\ref{B2}) can be employed to link $I_{l,l,n}(z)$ to
a lower-order integral $I_{l_0,l_0,n}(z)$, with $l_0<l$. When $n$ is even,
we may choose $l_0=0$. For odd $n\leq -3$ the recursion stops at a positive
value of $l$. As a consequence, one should choose $l_0=-\half n-\half>0$ in
that case. With these values of $l_0$ the solution of the recursion
relation reads for $l>l_0$:
\begin{eqnarray}
&&I_{l,l,n}(z)=\frac{(l_0-\half n+\half)_{l-l_0}}{(l_0+\half n+\threehalf)_{l-l_0}}I_{l_0,l_0,n}(z)\nonumber\\
&&-\frac{1}{2}z^{-n+1}\sum^{l-1}_{k=l_0+1}(2k+1)\frac{(k-\half
  n+\threehalf)_{l-k-1}}{(k+\half n+\half)_{l-k+1}}[h_k^{(1)}(z)]^2\nonumber\\
&&-\frac{1}{2}z^{-n+1}\frac{(l_0-\half n+\threehalf)_{l-l_0-1}}{(l_0+\half
  n+\threehalf)_{l-l_0}}[h_{l_0}^{(1)}(z)]^2\nonumber\\
&&-z^{-n+1}\frac{1}{2l+n+1}[h^{(1)}_l(z)]^2\, ,
\label{B4}
\end{eqnarray}
with Pochhammer symbols $(a)_n=a(a+1)\cdots(a+n-1)$. For $l=l_0+1$ the sum
drops out. As we shall see below, the sum can be evaluated in closed form
for all even $n\leq -2$ and for all odd $n\geq 1$. We shall consider in the
following the diagonal integrals with even $n=0, -2$ and with odd
$n=3,1,-1,-3,-5$, as these are needed in the main text.

For $n=0$ the expression (\ref{B4}) becomes, upon choosing $l_0=0$ and
evaluating the initial condition in terms of the exponential integral as
$I_{0,0,0}(z)=2iE_1(-2iz)+e^{2iz}/z$:
\begin{eqnarray}
&&I_{l,l,0}(z)=-\frac{2z}{2l+1}\sum_{k=0}^{l}[h^{(1)}_k(z)]^2+\frac{z}{2l+1}[h^{(1)}_l(z)]^2
\nonumber\\
&& +\frac{2i}{2l+1}E_1(-2iz)\, , 
\label{B5}
\end{eqnarray}
for all $l\geq 0$. The exponential integral can be rewritten \cite{NIST10}
in terms of sine and cosine integrals as
$E_1(-2iz)=-\mathrm{Ci}(2z)-i\mathrm{Si}(2z)+i\pi/2$.

For $n=-2$ one gets, again taking $l_0=0$ and inserting the (trivial)
result for $I_{0,0,-2}$:
\begin{eqnarray}
&&I_{l,l,-2}(z)=-2(2l+1)z^3\sum_{k=0}^l\frac{1}{(2k-1)(2k+3)}[h^{(1)}_k(z)]^2\nonumber\\
&&+\frac{z^3}{2l+3}[h^{(1)}_l(z)]^2+\frac{1}{2}(2l+1)(2z+i)e^{2iz}\, .
\label{B6}
\end{eqnarray}
The sum at the right-hand side can be evaluated with the help of
(\ref{C1}). In that way we arrive at the simpler result
\begin{eqnarray}
&&I_{l,l,-2}(z)=\frac{1}{2}z^3[h^{(1)}_l(z)]^2+\frac{1}{2}z^3[h^{(1)}_{l+1}(z)]^2\nonumber\\
&&-\frac{1}{2}(2l+1)z^2h^{(1)}_l(z)h^{(1)}_{l+1}(z)\, , 
\label{B7}
\end{eqnarray}
for $l\geq 0$.

Next we consider the diagonal integrals with odd values of $n$. For $n=3$
we choose $l_0=0$ in (\ref{B4}). For $l\geq 2$ the first and third terms at
the right-hand side drop out. As a result we get:
\begin{eqnarray}
&&I_{l,l,3}(z)=-\frac{1}{2(l-1)_4z^2}\sum_{k=1}^l (2k+1)k(k+1)[h^{(1)}_k(z)]^2\nonumber\\
&&+\frac{1}{2(l-1)z^2}[h^{(1)}_l(z)]^2\, , 
\label{B8}
\end{eqnarray}
for $l\geq 2$.  As before, the sum at the right-hand side can be evaluated
in explicit form in terms of $h^{(1)}_l$ and $h^{(1)}_{l+1}$, as is shown
in Appendix \ref{appendixc}. Upon substituting (\ref{C4}) for $p=1$ we
arrive at the expression:
\begin{eqnarray}
&&
\!\!\!\!\!\! I_{l,l,3}(z)=\left[\frac{z^2}{3(l-1)_4}+\frac{1}{6(l-1)l}+\frac{1}{2(l-1)z^2}\right][h^{(1)}_l(z)]^2\nonumber\\
&&+\left[\frac{z^2}{3(l-1)_4}+\frac{1}{6(l-1)(l+2)}\right][h^{(1)}_{l+1}(z)]^2\nonumber\\
&&-\left[\frac{2z}{3(l-1)l(l+2)}+\frac{1}{3(l-1)z}\right]h^{(1)}_l(z)h^{(1)}_{l+1}(z)\, ,
\label{B9}
\end{eqnarray}
for $l\geq 2$. This result is useless for $l=0,1$. These special cases can
be evaluated in terms of the exponential integral, as in (\ref{B5}).

For $n=1$ (and $l_0=0$) the first term at the right-hand side of (\ref{B4})
drops out for all $l\geq 1$. The special case $l=0$, which leads to an
exponential integral, has to be treated separately. For $l\geq 1$ one gets
an expression involving a sum that can be evaluated with the help of the
sum rule (\ref{C3}). The final result is
\begin{eqnarray}
&&I_{l,l,1}(z)=\left[\frac{z^2}{2l(l+1)}+\frac{1}{2l}\right][h^{(1)}_l(z)]^2\nonumber\\
&&+\frac{z^2}{2l(l+1)}[h^{(1)}_{l+1}(z)]^2-\frac{z}{l}h^{(1)}_l(z)h^{(1)}_{l+1}(z)\,
,
\label{B10}
\end{eqnarray}
for $l\geq 1$.  

For $n=-1$ we get from (\ref{B4}) (for $l_0=0$) upon inserting the initial
condition $I_{0,0,-1}(z)=E_1(-2iz)$:
\begin{eqnarray}
&&I_{l,l,-1}(z)=-\frac{1}{2}z^2\sum_{k=1}^l\frac{2k+1}{k(k+1)}[h^{(1)}_k(z)]^2\nonumber\\
&&+\frac{z^2}{2(l+1)}[h^{(1)}_l(z)]^2+E_1(-2iz)+\frac{1}{2}e^{2iz}\,
  ,  \label{B11}
\end{eqnarray}
for all $l\geq 0$. As it turns out, the sum cannot be simplified with the
help of one of the sum rules in Appendix \ref{appendixc}.

When considering the case $n=-3$ in (\ref{B4}) we have to choose $l_0=1$
and substitute the initial condition for $l=1$ that follows by a direct
evaluation as $I_{1,1,-3}(z)=E_1(-2iz)+(-\half iz+\fivefourth)e^{2iz}$. The
ensuing expression for $I_{l,l,-3}(z)$ contains a sum over squares of
spherical Hankel functions $h^{(1)}_k(z)$, with a coefficient
$(2k+1)/{(k-1)_4}$. This sum may be rewritten with the help of the identity
(\ref{C5}) for $p=1$. As a result we get for all $l\geq 0$:
\begin{eqnarray}
&&I_{l,l,-3}(z)=-\frac{1}{4}l(l+1)z^2\sum_{k=1}^l\frac{2k+1}{k(k+1)}[h^{(1)}_k(z)]^2\nonumber\\
&&+\frac{1}{4}z^4[h^{(1)}_l(z)]^2+\frac{1}{4}z^4[h^{(1)}_{l+1}(z)]^2\nonumber\\
&&-\frac{1}{2}lz^3h^{(1)}_l(z)h^{(1)}_{l+1}(z)+\frac{1}{2}l(l+1)E_1(-2iz)\nonumber\\
&&+\frac{1}{4}l(l+1)e^{2iz}\, ,
\label{B12}
\end{eqnarray} 
which contains the same sum as (\ref{B11}).

The last diagonal integral that we have to evaluate is the case with
$n=-5$. It follows from (\ref{B4}) by choosing $l_0=2$ and inserting the
initial condition for $l=2$, which has to be calculated by hand as
$I_{2,2,-5}(z)=9E_1(-2iz)+(\half iz^3-{\textstyle
  \frac{15}{4}}z^2-{\textstyle \frac{45}{4}}iz+{\textstyle
  \frac{117}{8}})e^{2iz}$. One encounters a sum over squares of spherical Hankel
functions with a coefficient containing $(k-2)_6$ in the denominator. Again
(\ref{C5}) (for $p=1,2$) is helpful in reducing this sum, so that we arrive
at the expression:
\begin{eqnarray}
&&I_{l,l,-5}(z)=-\frac{3}{16}(l-1)_4\,
z^2\sum_{k=1}^l\frac{2k+1}{k(k+1)}[h^{(1)}_k(z)]^2\nonumber\\
&&+\left[\frac{3}{16}l(l-1)+\frac{1}{8}z^2\right]z^4[h^{(1)}_l(z)]^2\nonumber\\
&&+\left[\frac{3}{16}(l-1)(l+2)+\frac{1}{8}z^2\right]z^4[h^{(1)}_{l+1}(z)]^2\nonumber\\  
&&-\left[\frac{3}{8}l(l+2)+\frac{1}{4}z^2\right](l-1)z^3h^{(1)}_l(z)h^{(1)}_{l+1}(z)\nonumber\\
&&+\frac{3}{8}(l-1)_4\, E_1(-2iz)+\frac{3}{16}(l-1)_4 e^{2iz}\, , \label{B13}
\end{eqnarray}
for all $l\geq 0$. The same sum as in (\ref{B11}) appears once again.

The off-diagonal integrals $I_{l,l+1,n}(z)$ with $n=-4,-2,-1,0,2$ follow
from the above diagonal integrals by employing (\ref{B3}).

To evaluate the dielectric decay-rate correction functions
$F_{\mathrm{d},\parallel}$ and $F_{\mathrm{d},\perp}$ we also need
information about indefinite integrals with spherical Hankel
functions and their complex conjugates in the integrand:
\begin{equation}
I'_{l_1,l_2,n}(z)=\int^z du\, u^{-n} h^{(1)}_{l_1}(u)\,
[h^{(1)}_{l_2}(u)]^\ast \, ,
\label{B14}
\end{equation}
for $z>0$, integer $n$ and non-negative integers $l_1,l_2$. Explicit
expressions for these integrals can be derived in a similar way as
above. The results are analogous, with some subtle differences. For $n=0$
and $l_1=l_2=l$ for instance, one finds on a par with (\ref{B5}) for all
$l\geq 0$:
 \begin{equation}
I'_{l,l,0}(z)=-\frac{2z}{2l+1}\sum_{k=0}^{l}|h^{(1)}_k(z)|^2+\frac{z}{2l+1}|h^{(1)}_l(z)|^2\,
. 
\label{B15}
\end{equation}
The term with the exponential integral is missing here, while the squares
of the spherical Hankel functions are replaced by the squares of their
moduli. We refrain from listing here the expressions for the other
$I'_{l_1,l_2,n}$ that are needed in the evaluation of
$F_{\mathrm{d},\parallel}$ and $F_{\mathrm{d},\perp}$.

\section{Sums of squares of spherical Hankel functions\label{appendixc}}
In Appendix \ref{appendixb} sums of squares of spherical Hankel functions
show up. Some of these may be evaluated as simple combinations of
$h^{(1)}_l$ and $h^{(1)}_{l+1}$. An example is the sum occurring in
(\ref{B6}). By induction with respect to $l$ one proves the following sum
rule for $l\geq 0$:
\begin{eqnarray}
&&\sum^l_{k=0}\frac{1}{(2k-1)(2k+3)}[h^{(1)}_k]^2=-\frac{1}{4(2l+3)}[h^{(1)}_l]^2\nonumber\\ 
&&-\frac{1}{4(2l+1)}[h^{(1)}_{l+1}]^2+\frac{1}{4z}h^{(1)}_lh^{(1)}_{l+1}
\nonumber\\
&&+\left(\frac{1}{2z^2}+\frac{i}{4z^3}\right)e^{2iz}\, ,
\label{C1}
\end{eqnarray}
where we omitted the argument $z$ of the spherical Hankel functions.  It
turns out that this sum rule is the first in a hier\-archy of sum rules
with an increasing number of factors in the denominator of the coefficient
in the summand. In fact, one may prove for all $p\geq 0$ and all $l\geq 0$:
\begin{eqnarray}
&&\sum_{k=0}^l\frac{2k+1}{(k-p-\half)_{2p+3}}[h^{(1)}_k]^2=\nonumber\\
&&
\!\!\!\!\!\!\!\! =-\frac{1}{2}\left[\sum_{k=0}^p\frac{(p-k+1)_k}
{(p-k+\frac{1}{2})_{k+1}(l-p+k+\frac{3}{2})_{2p-2k+1}}\frac{1}{z^{2k}}\right][h^{(1)}_l]^2\nonumber\\
&&\!\!\!\!\!\!\!\! -\frac{1}{2}\left[\sum_{k=0}^p
\frac{(p-k+1)_k}{(p-k+\frac{1}{2})_{k+1}(l-p+k+\frac{1}{2})_{2p-2k+1}}\frac{1}{z^{2k}}\right][h^{(1)}_{l+1}]^2\nonumber\\
&&\!\!\!\!\!\!\!\! +\left[\sum_{k=0}^p
\frac{(p-k+1)_k}{(p-k+\frac{1}{2})_{k+1}(l-p+k+\frac{3}{2})_{2p-2k}}\frac{1}{z^{2k+1}}\right]h^{(1)}_lh^{(1)}_{l+1}\nonumber\\
&&+R_{1,p}(z)\, .
\label{C2}
\end{eqnarray}
The last term, which is independent of $l$, follows by taking $l=0$ on both
sides of the identity. For small values of $p$ the polynomials in $1/z$ are
of low degree, so that this identity yields an efficient way of evaluating
the sum, in particular for higher $l$.

A second hierarchy of sum rules, with an increasing number of factors in
the numerator of the summand, can be established as well. The first sum
rule in this hierarchy reads
\begin{eqnarray}
&&\sum_{k=0}^l(2k+1)[h^{(1)}_k]^2=\nonumber\\
&&=-z^2[h^{(1)}_l]^2-z^2[h^{(1)}_{l+1}]^2+2(l+1)z h^{(1)}_lh^{(1)}_{l+1}\,
, 
\label{C3}
\end{eqnarray}
for $l\geq 0$. The complete hierarchy reads for all
$p\geq 0$ and $l\geq p$:
\begin{eqnarray}
&&\sum_{k=p}^l(2k+1)(k-p+1)_{2p} [h^{(1)}_k]^2=\nonumber\\
&&\!\!\!\!\!\!\!\! =-\frac{1}{2}\left[\sum_{k=0}^p\frac{(p-k+1)_k(l-p+k+2)_{2p-2k}}{(p-k+\frac{1}{2})_{k+1}}z^{2k+2}\right][h^{(1)}_l]^2\nonumber\\
&&\!\!\!\!\!\!\!\! -\frac{1}{2}\left[\sum_{k=0}^p\frac{(p-k+1)_k(l-p+k+1)_{2p-2k}}{(p-k+\frac{1}{2})_{k+1}}z^{2k+2}\right][h^{(1)}_{l+1}]^2\nonumber\\
&&\!\!\!\!\!\!\!\! +\left[\sum_{k=0}^p\frac{(p-k+1)_k(l-p+k+1)_{2p-2k+1}}{(p-k+\frac{1}{2})_{k+1}}z^{2k+1}\right] h^{(1)}_l h^{(1)}_{l+1}\, .
\nonumber\\
&&
\label{C4}
\end{eqnarray} 
A remainder function independent of $l$ does not occur here.

Apart from these sum rules one may prove several hierarchies of identities
relating sums of a similar type as above. A first one of these reads as
follows
\begin{eqnarray}
&&\!\!\!\!\!\!\!\! z^2\sum_{k=p+1}^l\frac{2k+1}{(k-p)_{2p+2}}[h^{(1)}_k]^2=\frac{2p-1}{2p}
\sum_{k=p}^l\frac{2k+1}{(k-p+1)_{2p}}[h^{(1)}_k]^2\nonumber\\
&&-\frac{1}{2p(l-p+2)_{2p}}z^2[h^{(1)}_l]^2-\frac{1}{2p(l-p+1)_{2p}}z^2[h^{(1)}_{l+1}]^2\nonumber\\
&&+\frac{1}{p(l-p+2)_{2p-1}}z h^{(1)}_l
h^{(1)}_{l+1}+R_{2,p}(z)\, ,
\label{C5}
\end{eqnarray}
for $p\geq 1$ and $l\geq p$. The remainder function $R_{2,p}(z)$ follows by
putting $l=p$, and discarding the left-hand side. In Appendix
\ref{appendixb} this identity has been employed for $p=1$ and $p=2$, in
order to reduce several sums over squares of spherical Hankel functions
with complicated coefficients to sums with simpler coefficients.

Yet another hierarchy of identities relates sums with coefficients
containing an increasing number of factors in the numerator:
\begin{eqnarray}
&&\sum_{k=0}^l (2k+1)(k-p+\half)_{2p+1}[h^{(1)}_k]^2=\nonumber\\
&&=\frac{2p+1}{2(p+1)}z^2\sum_{k=0}^l (2k+1)(k-p+\threehalf)_{2p-1}[h^{(1)}_k]^2\nonumber\\
&&-\frac{(l-p+\frac{3}{2})_{2p+1}}{2(p+1)} z^2[h^{(1)}_l]^2-\frac{(l-p+\frac{1}{2})_{2p+1}}{2(p+1)}z^2[h^{(1)}_{l+1}]^2\nonumber\\
&&+\frac{(l-p+\frac{1}{2})_{2p+2}}{p+1}zh^{(1)}_l h^{(1)}_{l+1}+R_{3,p}(z) \, , 
\label{C6}
\end{eqnarray} 
for $p\geq 0$ and $l\geq 0$. For $p=0$ the Pochhammer symbol
$(k+\frac{3}{2})_{-1}$ in the sum at the right-hand side should be read as
$1/(k+\frac{1}{2})$, as follows by using the alternative form
$(a)_n=\Gamma(a+n)/\Gamma(a)$ \cite{NIST10}. The remainder function
$R_{3,p}(z)$ is independent of $l$ and follows by taking $l=0$.

Similar sum rules may be established for sums over squares of the modulus
$|h^{(1)}_l(z)|$ of a spherical Hankel function (with a real argument $z$),
or, even more generally, for sums over products $f_l(z)g_l(z)$, with $f_l$
and $g_l$ equal to $j_l$, $y_l$, $h_l^{(1)}$ or $h_l^{(2)}$, independently. For
instance, one may prove a sum rule like (\ref{C2}), with all
$[h^{(1)}_k]^2$ replaced by $|h^{(1)}_k|^2$ and the product $h^{(1)}_l
h^{(1)}_{l+1}$ replaced by its real part $\half [h^{(1)}_l
h^{(2)}_{l+1}+h^{(1)}_{l+1} h^{(2)}_l]$ (for real arguments). The remainder
function is found to be 0 in that case. The analogues of the sum rules
(\ref{C2}), (\ref{C4}) and (\ref{C5}) for the modulus of the spherical
Hankel functions are useful in deriving suitable expressions for the
integrals $I'_{l_1,l_2,n}$ in Appendix \ref{appendixb}.

The above sum rules, with squares $[j_k(z)]^2$ instead of
$[h^{(1)}_k(z)]^2$, may be employed to derive identities for infinite sums
that have been used in Section \ref{sectionevaluation}. The analogue of
(\ref{C4}) for spherical Bessel functions contains a remainder function $p!
z^{2p}/(\threehalf)_p$. Upon taking the limit $l\rightarrow \infty$, the other
terms at the right-hand side drop out, so that one is left with the
equality
\begin{equation}
\sum_{k=p}^\infty (2k+1)(k-p+1)_{2p}
[j_k(z)]^2=\frac{p!}{(\threehalf)_p}z^{2p}\, ,
\label{C7}
\end{equation}
for $p\geq 0$. For $p=0$ this sum rule is well-known \cite{NIST10}, while
for general $p\geq 0$ it agrees with an identity in \cite{L69}. Likewise,
the analogue of (\ref{C6}) for $[j_k(z)]^2$ contains the remainder function
$-(-p-\half)_{2p+1}[\cos(2z)+(p+\half) \sin(2z)/z]/(2(p+1))$. In the limit
of infinite $l$ the second, third and fourth terms at the right-hand side
drop out, so that a simple recursion relation connecting infinite sums is
obtained. It may be combined with the initial condition \cite{NIST10}
\begin{equation}
\sum_{k=0}^\infty [j_k(z)]^2=\frac{\mathrm{Si}(2z)}{2z} 
\label{C8}
\end{equation}
to derive expressions for all sums of the form $\sum_{k=0}^\infty
(2k+1)(k-p+\half)_{2p+1} [j_k(z)]^2$ with $p\geq 0$. In particular, one gets
for $p=0$:
\begin{equation}
\sum_{k=0}^\infty (2k+1)^2 [j_k(z)]^2=z \mathrm{Si}(2z)
+\frac{\sin(2z)}{4z}+ \frac{1}{2}\cos(2z)\, ,
\label{C9}
\end{equation}
which is consistent with an identity involving the generalized
hypergeometric function $\mbox{}_1F_2(\half;\fivehalf,\fivehalf;-z^2)$ in
\cite{L69}. Finally, an infinite sum rule with alternating signs:
\begin{equation}
\sum_{k=p}^\infty (-1)^k (2k+1)(k-p+1)_{2p} [j_k(z)]^2=(-1)^p p!  z^p
j_p(2z)
\label{C10}
\end{equation}
has been employed in Section \ref{sectionevaluation} for $p=0,1$. For $p=0$
it is well-known \cite{NIST10}, whereas for general $p\geq 0$ it can be found
in \cite{L69}.

\twocolumngrid

\end{document}